\pdfoutput=1
\documentclass[10pt,twocolumn]{article}
\usepackage[a4paper, margin=1.5cm]{geometry}
\usepackage{indentfirst}
\usepackage[utf8]{inputenc}
\usepackage[english]{babel}
\usepackage{csquotes}
\usepackage{amsmath}
\usepackage{amsfonts}
\usepackage{amssymb}
\usepackage[per-mode=symbol]{siunitx}
\DeclareSIUnit\angstrom{\text{Å}}
\usepackage{bm}                       
\usepackage{mathpazo}                 
\usepackage{microtype}                
\usepackage{tikz}
\usepackage{graphicx}
\usepackage{float}                    
\usepackage[export]{adjustbox}        
\usepackage{balance}                  
\usepackage[font=small,labelfont=bf]{caption}
\usepackage{subfig}
\usepackage{ifthen}                   
\usepackage{booktabs}                 
\usepackage{multirow}                 
\usepackage[symbol]{footmisc}         
\usepackage[version=4]{mhchem}        

\usepackage{hyperref}
\hypersetup{colorlinks=true, linkcolor=blue, citecolor=blue, urlcolor=blue} 

\usepackage{xcolor}
\definecolor{blue1}{RGB}{ 7,  47,  95}
\definecolor{blue2}{RGB}{18,  97, 160}
\definecolor{blue3}{RGB}{56, 149, 211}
\definecolor{red}{RGB}{210, 0, 0}

\usepackage{titlesec}                 
\titlelabel{\thetitle.\ }             
\titleformat*{\section}{\color{blue1}\scshape\bfseries\centering\large}
\titleformat*{\subsection}{\color{blue2}\normalfont\itshape\large}
\titleformat*{\subsubsection}{\color{blue3}\normalfont\itshape}
\titleformat{\paragraph}[runin]{\color{blue3}\normalfont\itshape}{}{0em}{}[~--]
\titlespacing{\paragraph}{0em}{0em}{0.3em}

\usepackage[backend=bibtex,date=year,doi=false,eprint=false,giveninits=true,isbn=false,maxnames=1,maxbibnames=99,sorting=none,style=nature]{biblatex}
\AtEveryBibitem{
	\clearlist{language}
}
\newlength{\spc} 
\let\footnoteorig\footnote
\renewcommand{\footnote}[2]{
	\ifthenelse{\equal{#2}{,}\OR\equal{#2}{.}}{%
		\settowidth{\spc}{#2}
		\addtolength{\spc}{-1.8\spc}
		#2
		\hspace*{\spc}
		\footnoteorig{#1}
	}{%
		\footnoteorig{#1}%
		\ifthenelse{\NOT\equal{#2}{;}\AND\NOT\equal{#2}{:}}{\ }{}%
		#2%
	}%
} 

\renewcommand{\textcite}[1]{\citeauthor{#1}\hspace*{-0.15em}\supercite{#1}}
\renewcommand{\cite}[2]{
	\ifthenelse{\equal{#2}{,}\OR\equal{#2}{.}}{%
		\settowidth{\spc}{#2}
		\addtolength{\spc}{-1.8\spc}
		#2
		\hspace*{\spc}
		\supercite{#1}
	}{%
		\supercite{#1}%
		\ifthenelse{\NOT\equal{#2}{;}\AND\NOT\equal{#2}{:}}{\ }{}%
		#2%
	}%
}

\newcommand{\snspace}[2][0.45em]{
	\hspace*{-#1}#2\hspace{0.2em}}

\begin{document}

\twocolumn[
	\begin{@twocolumnfalse}

		\begin{center}
			\textbf{\color{blue1}\large Adhesive wear with a coarse-grained discrete element model}\\
			\vspace{1em}
			Son Pham-Ba\footnotemark\hspace*{-0.3em},\hspace{0.1em} Jean-François Molinari\\\vspace{0.5em}
			\textit{\footnotesize Institute of Civil Engineering, Institute of Materials Science and Engineering,\\\vspace{-0.2em}
			École polytechnique fédérale de Lausanne (EPFL), CH 1015 Lausanne, Switzerland}
		\end{center}


		\begin{center}
			\parbox{14cm}{\small
				\setlength\parindent{1em}The use of molecular dynamics (MD) simulations has led to promising results to unravel the atomistic origins of adhesive wear, and in particular for the onset of wear at nanoscale surface asperities.  However, MD simulations come with a high computational cost and offer access to only a narrow window of time and length scales. We propose here to resort to the discrete element method (DEM) to mitigate the computational cost. Using DEM particles with contact and cohesive forces, we reproduce the key mechanisms observed with MD, while having particle diameters and system sizes an order of magnitude higher than with MD. The pairwise forces are tuned to obtain a solid with reasonably approximated elastic and fracture properties. The simulations of single asperity wear performed with MD are successfully reproduced with DEM using a range particle sizes, validating the coarse-graining procedure. More complex simulations should allow the study of wear particles and the evolution of worn surfaces in an adhesive wear context, while reaching scales inaccessible to MD.

				\vspace{1em}
			}
		\end{center}

		\vspace{1em}

	\end{@twocolumnfalse}
]

\footnotetext{Corresponding author. E-mail address: \href{mailto:son.phamba@epfl.ch}{son.phamba@epfl.ch}}

\section{Introduction}

Wear is a phenomenon occurring at a wide range of length scales. It manifests itself at sliding geological faults\cite{scholzWearGougeFormation1987,rechesGougeFormationDynamic2005}, creating a third-body layer (also called gouge) which directly influences the frictional properties of the interface\cite{biegelFrictionalPropertiesSimulated1989,mairInfluenceGrainCharacteristics2002}. The presence of gouge has implications on seismic events\cite{mizoguchiReconstructionSeismicFaulting2007}. At our everyday scale, the wear of car brakes or of tyres on roads is responsible for more than half of traffic-related air pollution\cite{grigoratosBrakeWearParticle2015}, and the wear of tyres in particular has a significant impact on the quantity of microplastics in the environment\cite{koleWearTearTyres2017}. At the smaller scales, the presence of wear is less spectacular but still existing, for example in nanoelectromechanical systems (NEMS) that could be used for high-density data storage\cite{vettigerUltrahighDensityHighdatarate1999,bhushanPlatinumcoatedProbesSliding2007}. At this scale, wear is studied using atomic force microscopy probes\cite{vahdatMechanicsInteractionAtomicScale2013,aghababaeiCriticalLengthScale2016} having undamaged tip radii of the order of \SI{20}{nm}.

A better control of wear and its consequences calls for a fundamental understanding, which can be achieved by means of numerical simulations. A sensible approach is to investigate wear at the level of asperities, which is amenable to both MD simulations and AFM experiments. Molecular dynamics (MD) simulations were used to study systems ranging from the wear of a single asperity\cite{aghababaeiCriticalLengthScale2016,aghababaeiDebrislevelOriginsAdhesive2017,zhaoAdhesiveWearLaw2020,aghababaeiMicromechanicsMaterialDetachment2021} to the growth of multiple third-body particles trapped between two sliding surfaces in three dimensions\cite{aghababaeiAsperityLevelOriginsTransition2018,milaneseEmergenceSelfaffineSurfaces2019a,brinkEffectWearParticles2022}. The latter simulations show that the rolling particles grow into rolling cylinders and merge together into a gouge layer, with a noticeable effect on the macroscopic tangential force resisting the sliding motion. The formation of rolling cylinders and of a gouge layer from third-body particles is also observed experimentally\cite{chenPowderRollingMechanism2017,pham-baCreationEvolutionRoughness2021}, showcasing the importance of modelling multiple third-body particles and their interactions during sliding. However, the largest adhesive wear MD simulations (\emph{e.g.} by \textcite{brinkEffectWearParticles2022}) start to reach a computational barrier, having around \SI{35000000} atoms per simulation. Due to the very small size of the atoms simulated in MD (order of $\SI{1}{\angstrom}$), the simulations are limited both in space and in time (the time step must also be small, of the order of $\SI{1}{ps}$).

Having possibly reached the maximum capabilities of MD regarding scale, other methods must be used to further increase the size of the simulated worn systems and explore the effects of collective mechanisms, as these mechanisms are ultimately responsible for the macroscopic wear response. Despite the small scale disadvantage, the benefit of MD simulations is to seamlessly model, with simple force potentials, particle rearrangements such as fracture and mixing of materials, which is much more challenging to achieve in continuum methods like the widely used finite element method. To preserve the advantages of MD while going to larger scales, we propose to resort to the discrete element method (DEM) to coarse-grain the particle interactions. DEM is a popular method to model the flow of granular media in general, and in particular rocks and gouge\cite{pandeNumericalMethodsRock1990} in geomechanics. It is also used to model third-body rheology\cite{fillotSimulationWearMass2005,fillotModellingThirdBody2007}, while including other physical effects\cite{renoufNumericalTribologyDry2011} (\emph{e.g.} thermal). The discrete particles are commonly modeled with breakable bonds\cite{cundallComputerModelSimulating1971} to represent crushable material. The particles can also be modeled with adhesive/cohesive forces such as JKR\cite{johnsonSurfaceEnergyContact1971}, but the resulting systems made of many particles have elastic properties which are dependent on a confinement pressure and are challenging to predict\cite{chengElasticWavePropagation2020,voisin--leprinceEnergybasedCouplingApproach2022}. The DEM is also (less commonly) used to model continuum media. However, the link between the interaction properties between the particles and the macroscopic elastic properties of the assembly of particles is not straightforward when simple spring forces are used between the particles\cite{hentzIdentificationValidationDiscrete2004,jerierNormalContactRough2012}. To exactly match some desired elastic properties, the forces between particles must take into account the neighborhood of each particle\cite{celiguetaAccurateModellingElastic2017}, making their formulation more complex. Capturing the Poisson's effect is also shown to be challenging using only linear spring forces. To the best of our knowledge, damage is most often irreversible in DE models modeling fracture, and the reattachment of matter is not considered.

To perform the same kind of adhesive wear simulation as with MD but using DEM, we aim to model a solid with known elastic and fracture properties, and the fracture process must be reversible to capture the growth of rolling third-body particles in a sheared interface, which involves reattachment of matter due to adhesive forces. We formulate a DE method suited for this problem, with relatively simple pair forces to remain computationally inexpensive. The interaction forces have a repulsive part and a reversible cohesive part, inspired from the most simple MD pair forces (\emph{e.g.} Lennard-Jones). In Section~\ref{sec:method}, we present our formulation of pair forces and explain how its parameters are tuned to match the elastic and fracture properties with an assembly of many particles. Then, in Section~\ref{sec:validation}, the model and the choice of its parameters are validated using simple patch tests. Finally, we show in Section~\ref{sec:wear} an example of application of the adhesive wear of a single junction between two sliding surfaces, similar to what was done using MD\cite{aghababaeiCriticalLengthScale2016}.

\section{Method}\label{sec:method}

\begin{table}
	\centering
	\caption{List of symbols used for lengths and sizes}
	\label{tab:symbols}
	\vspace{-0.5em}
    \small
\begin{tabular}{ll}
    \toprule
    \textbf{Symbol} & \textbf{Description} \\
    \midrule
    $d$ & Particle diameter \\
    $d_\text{min}$ & Minimum acceptable particle diameter \\
    $d_\text{c}$ & Critical particle diameter \\
    $d_0$ & Average particle size \\
    $d_\text{s}$, $d_\text{l}$ & Smallest and largest bounds in size distribution \\
    $d^*$ & Critical material length scale \\
    $D$ & Junction size \\
    $r_i$, $r_j$ & Particle radii \\
    $\delta_\text{N}$, $\delta_\text{T}$ & Normal and tangential particle separations \\
    $\delta_\text{e}$ & Elastic separation \\
    $\delta_\text{f}$ & Fracture separation \\
    \bottomrule
\end{tabular}
\end{table}

The three dimensional physical system is discretized into many spherical particles (each identified by an index $i$) of radius $r_i$ and density $\rho$ (see Table~\ref{tab:symbols} for a list of used symbols). Forces of interaction are acting between every pair of distinct particles, and the particles' velocities and positions are updated accordingly using the semi-implicit Euler method (also called symplectic Euler) with a time step $\Delta t$. For any given particle, the integration scheme between steps $n$ and $n + 1$ is
\begin{subequations}\label{eq:euler}
\begin{align}
	\bm{v}_{n + 1} &= \bm{v}_n + \frac{\bm{F}_n}{m} \, \Delta t \,, \\
	\bm{x}_{n + 1} &= \bm{x}_n + \bm{v}_{n + 1} \, \Delta t \,, \\
	\bm{\omega}_{n + 1} &= \bm{\omega}_n + \bm{T}_n \, \Delta t \,,
\end{align}
\end{subequations}
where $\bm{v_n}$, $\bm{x_n}$ and $\bm{\omega}_n$ are respectively the velocity, the position and the angular velocity of the particle, $\bm{F}_n$ and $\bm{T}_n$ are the force and the torque acting on the particle, and $m$ is the mass of the particle. The presence of $\bm{v}_{n+1}$ in the expression for $\bm{x}_{n+1}$ is what makes the scheme semi-implicit, without actually requiring to solve any implicit equation to perform a step. This integration scheme is the one currently implemented in the open-source software LAMMPS\cite{plimptonFastParallelAlgorithms1995} we are using.

\begin{figure}
	\centering
	\includegraphics{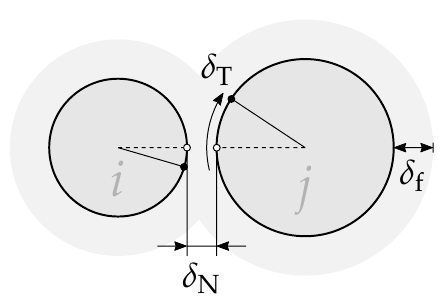}
	\caption{Two interacting particles}
	\label{fig:dem_deltas}
\end{figure}

For each pair of particles $(i, j)$, we define the normal distance $\delta_\text{N}$ between their surfaces, and the tangential sliding distance $\delta_\text{T}$ (see Figure~\ref{fig:dem_deltas}). The normal distance is simply equal to $d_{ij} - r_i - r_j$, with $d_{ij}$ being the distance between the particles' centers. We have $\delta_\text{N} = 0$ when the particles are touching and $\delta_\text{N} < 0$ when they are interpenetrated. The tangential sliding distance $\delta_\text{T}$ is only defined when the particles are within their range of interaction (shaded area around particles in Figure~\ref{fig:dem_deltas}, more details later). $\delta_\text{T}$ is equal to $0$ when the particles start interacting, and is updated using the relative rolling velocity\cite{wangRollingSliding3D2015}. $\delta_\text{T}$ is always positive (or equal to zero).

\subsection{Forces between particles}

The force $\bm{F}$ acting between a given pair of particles is the sum of a normal component $F_\text{N}$, a tangential component $F_\text{T}$, and velocity damping forces:
\begin{equation}
	\bm{F} = -(F_\text{N} + c_\text{N}v_\text{N}) \bm{n}_\text{N} - (F_\text{T} + c_\text{T}v_\text{T}) \bm{n}_\text{T} \,,
\end{equation}
where $\bm{n}_\text{N}$ and $\bm{n}_\text{T}$ are the unit vectors pointing respectively in the normal and tangential directions, the latter being computed using the evolution of the rolling velocity\cite{wangRollingSliding3D2015}, $v_\text{N}$ and $v_\text{T}$ are the corresponding relative velocities at the point of interaction, and $c_\text{N}$ and $c_\text{T}$ are damping factors. Since the total force $\bm{F}$ acts on the surface of the particles, it also induces torques $\bm{T}$ (when seen from the centers of the particles), computed directly from $\bm{F}$ and the appropriate moment arms. In order to model an elastic solid with the discrete particles, cohesive forces between particles are needed in addition to the usually modelled repulsive contact forces, all of which are defined thereafter.

\paragraph{Normal force}

\begin{figure}
	\centering
	\includegraphics{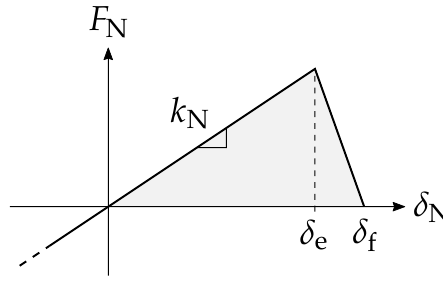}
	\caption[Normal force between two particles as a function of inter-particular distance $\delta_\text{N}$]{Normal force between two particles as a function of inter-particular distance $\delta_\text{N}$. There is interpenetration when $\delta_\text{N} < 0$. The force has a cohesive part when $\delta_\text{N} > 0$.}
	\label{fig:dem_forces}
\end{figure}

The normal component of the pairwise force depends on the inter-particular distance $\delta_\text{N}$ and has the profile shown in Figure~\ref{fig:dem_forces}. When the particles are interpenetrating ($\delta_\text{N} \leqslant 0$), they feel a Hookean repulsive force $F_\text{N} = k_\text{N} \delta_\text{N}$, where $k_\text{N}$ is the normal stiffness. When the particles are not touching ($\delta_\text{N} > 0$), we model a cohesive force by keeping the Hookean force up to a separation $\delta_\text{e}$, until which the interaction between the particles is \emph{elastic} (hence the subscript letter `e' in $\delta_\text{e}$). The fracture process is modelled by a linear weakening zone between the elastic separation $\delta_\text{e}$ and a \emph{fracture} separation $\delta_\text{f}$ (see Figures~\ref{fig:dem_deltas} and \ref{fig:dem_forces}). When $\delta_\text{N} > \delta_\text{f}$, the particles are not interacting, and the total force is zero. The full expression of the normal force is
\begin{equation}\label{eq:FN}
	F_\text{N} = \begin{cases}
		k_\text{N} \delta_\text{N} & \text{if } \delta_\text{N} \leqslant \delta_\text{e} \,, \\
		\displaystyle -\frac{k_\text{N} \delta_\text{e}}{\delta_\text{f} - \delta_\text{e}} (\delta_\text{N} - \delta_\text{f}) & \text{if } \delta_\text{e} < \delta_\text{N} \leqslant \delta_\text{f} \,, \\
		0 & \text{otherwise.}
	\end{cases}
\end{equation}
The value of the normal force $F_\text{N}$ is independent of the history of $\delta_\text{N}$. Therefore, the fracture process is fully reversible, and particles can create or recreate new `bonds´ with neighboring particles. We make the simplifying assumption that newly created bonds have the same properties (stiffness, strength) as previously existing bonds, which is not always the case in reality, as some phenomena can weaken the reattachment (\emph{e.g.} surface roughness or oxidation).

\paragraph{Tangential force}

\begin{figure}
	\centering
	\includegraphics{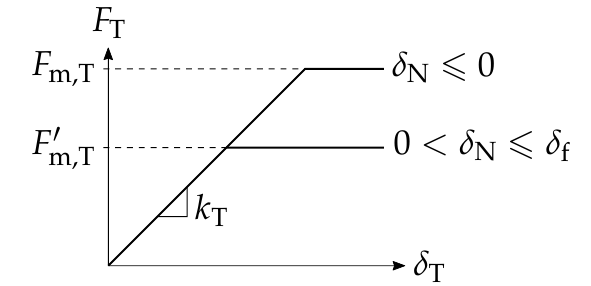}
	\caption[Tangential force between two particles as a function of sliding distance $\delta_\text{T}$]{Tangential force between two particles as a function of sliding distance $\delta_\text{T}$. When the particles are in contact ($\delta_\text{N} \leqslant 0$), the tangential force is bounded by $F_\text{m,T}$, whereas when they are not touching but still in their range of interaction ($0 < \delta_\text{T} \leqslant \delta_\text{f}$), the force is bounded by $F'_\text{m,T}$. The corrected bound force $F'_\text{m,T}$ is equal to $F_\text{m,T}$ when $\delta_\text{N} = 0$ and decreases linearly down to 0 when $\delta_\text{N} = \delta_\text{f}$.}
	\label{fig:dem_forces_T}
\end{figure}

The tangential component of the force depends on the sliding distance $\delta_\text{T}$ and has the profile shown in Figure~\ref{fig:dem_forces_T}. When the particles are interpenetrating, the force has the expression $F_\text{T} = k_\text{T} \delta_\text{T}$ up to a maximum value of $F_\text{m,T}$. When the particles are not touching but still in their range of interaction ($0 < \delta_\text{N} \leqslant \delta_\text{f}$), the maximum reachable force $F_\text{m,T}$ is decreased from its original value (at $\delta_\text{N} = 0$) toward zero at $\delta_\text{N} = \delta_\text{f}$. The full expressions of the tangential force $F_\text{T}$ and the rescaled maximal tangential force $F_\text{m,T}'$ are
\begin{align}
	&F_\text{T} = \min(k_\text{T} \delta_\text{T}, F_\text{m,T}') \,, \\
	&F_\text{m,T}' = \min\left(\frac{\delta_\text{f} - \delta_\text{N}}{\delta_\text{f}}, 1\right) F_\text{m,T} \,.
\end{align}

\paragraph{List of force parameters}

In summary, there are 7 parameters, listed in Table~\ref{tab:force_params}.

\begin{table}[H]
	\centering
	\caption{List of force parameters}
	\label{tab:force_params}
	\vspace{-0.5em}
	\small
\begin{tabular}{lc}
	\toprule
	\textbf{Name}            & \textbf{Symbol} \\
	\midrule
	Normal stiffness         & $k_\text{N}$ \\
	Tangential stiffness     & $k_\text{T}$ \\
	Elastic separation       & $\delta_\text{e}$ \\
	Fracture separation      & $\delta_\text{f}$ \\
	Maximum tangential force & $F_\text{m,T}$ \\
	Normal damping           & $c_\text{N}$ \\
	Tangential damping       & $c_\text{T}$ \\
	\bottomrule
\end{tabular}

\end{table}

\paragraph{Simulations' time step}

In equations \eqref{eq:euler}, we stated the integration scheme used with our model, which depends on a time step $\Delta t$. The scheme is different from the more commonly used central difference scheme\cite{cundallDiscreteNumericalModel1979}. In order for the simulation to be numerically stable, we define a critical time step for our particular integration scheme (derived in Appendix~\ref{apx:dt_crit})
\begin{equation}
	\Delta t_\text{c} = \sqrt{\frac{2m}{k_\text{N}}} \,, \label{eq:Dtc}
\end{equation}
which is the maximum time step at which a simulation comprised of two particles in contact in the linear Hookean range ($\delta_\text{N} \leqslant \delta_\text{e}$) remains stable. For a system with many particles, we typically choose a time step being a fraction of the critical time step.

\subsection{Matching macroscopic material properties}

For each pair $(i, j)$ of particles, the parameters of the interaction forces can be tuned such that the assembly of many particles exhibit the desired mechanical properties. The choice of the normal and tangential stiffnesses $k_\text{N}$ and $k_\text{T}$ determines the macroscopic Young's modulus $E$ and the Poisson's ratio $\nu$. The elastic domain extends up to the interparticular distance $\delta_\text{N} = \delta_\text{e}$, so $\delta_\text{e}$ controls the macroscopic tensile strength $\sigma_\text{m,N}$. In the same manner, the maximum tangential force $F_\text{m,T}$ controls the macroscopic shear strength $\sigma_\text{m,T}$\footnote{This is a simplified view. In fact, in an assembly if many particles, a tensile stress will displace the particles both in the normal and in the tangential directions relative to each other, so that both the tensile and the shear strength will contribute to the actual strength of the assembly. The same is true for a shear motion.}. The interaction distance $\delta_\text{f}$ defines the surface energy $\gamma$, which is linked to the fracture energy (shaded area in Figure~\ref{fig:dem_forces} under the force-displacement curve). Finally, the damping factors $c_\text{N}$ and $c_\text{T}$ influence the restitution coefficient $\eta$, which is the ratio between final and initial relative velocities when two particles collide. The list of material properties needed to fully determine the force parameters is given in Table~\ref{tab:material_params}.

\begin{table}[H]
	\centering
	\caption{List of target material properties}
	\label{tab:material_params}
	\vspace{-0.5em}
	\small
\begin{tabular}{lc}
	\toprule
	\textbf{Name}            & \textbf{Symbol} \\
	\midrule
	Young's modulus         & $E$ \\
	Poisson's ratio         & $\nu$ \\
	Tensile strength        & $\sigma_\text{m,N}$ \\
	Shear strength          & $\sigma_\text{m,T}$ \\
	Surface energy          & $\gamma$ \\
	Restitution coefficient & $\eta$ \\
	Density                 & $\rho$ \\
	\bottomrule
\end{tabular}

\end{table}

Choosing the right force properties to obtain some desired macroscopic material properties is a knowingly challenging task for this kind of DEM model. As a first guess in the process of calibrating the parameters, they can be expressed in term of material properties (the derivation of the expressions is given in Appendix~\ref{apx:force_param}):
\begin{align}
	&k_\text{N} = \frac{A_\text{N}E}{r_i + r_j} \,, \label{eq:kN} \\
	&k_\text{T} = \frac{A_\text{T}E}{r_i + r_j} \,, \label{eq:kT} \\
	&\delta_\text{e} = \frac{(r_i + r_j) \sigma_\text{m,N}}{E} \,, \\
	&\delta_\text{f} = \begin{cases}
		\displaystyle \frac{4\gamma}{\sigma_\text{m,N}} & \text{if } r_i + r_j \leqslant d_\text{c} \,, \\
		\displaystyle \delta_\text{e} \left( \frac{r_i + r_j}{d_\text{c}} \right)^{-s} & \text{otherwise,}
		\end{cases} \label{eq:df} \\
	&F_\text{m,T} = A_\text{T} \, \sigma_\text{m,T} \,, \\
	&c_\text{N} = \frac{2(1 - \eta)}{\pi} \sqrt{k_\text{N}m_\text{eff}} \,, \label{eq:cN} \\
	&c_\text{T} = \frac{2(1 - \eta)}{\pi} \sqrt{k_\text{T}m_\text{eff}} \,. \label{eq:cT}
\end{align}
$A_\text{N}$ and $A_\text{T}$ are effective contact cross section between the interacting particles, chosen to have the macroscopic properties correctly scaled and to balance between normal and tangential forces (see equations \eqref{eq:kN} and \eqref{eq:kT}) in order to obtain the correct Poisson's ratio. The effective cross sections are dependent on an effective particle radius
\begin{equation}
	r_\text{eff} = \min(r_i, r_j)
\end{equation}
and are defined as
\begin{align}
	A_\text{N} &= \sqrt{2} r_\text{eff}^2 \frac{1}{1 - 2\nu} \,, \\
	A_\text{T} &= \sqrt{2} r_\text{eff}^2 \frac{1 - 4\nu}{(1 - 2\nu)(1 + \nu)} \,.
\end{align}
From these expressions, we note that the target Poisson's ratio can only take values up to $\nu = 1/4$ and values between $1/4$ and $1/2$ cannot be modeled. In \eqref{eq:cN} and \eqref{eq:cT}, $m_\text{eff}$ is the effective mass of the oscillatory system comprised of the two interacting particles:
\begin{equation}
	m_\text{eff} = \frac{m_i m_j}{m_i + m_j} \,.
\end{equation}
In the expression of the fracture distance $\delta_\text{f}$ \eqref{eq:df}, $d_\text{c}$ is a critical diameter, defined as
\begin{equation}
	d_\text{c} = \frac{4 \gamma E}{\sigma_\text{m,N}^2} \,, \label{eq:dc}
\end{equation}
and $s \geqslant 0$ is a scaling parameter.

\paragraph{Effect of particle size}

Note that nearly all parameters depend on the size of the particles. In particular, $\delta_\text{f}$ takes two different expressions depending on the particles' sizes. When the particles are smaller than the critical diameter ($r_i + r_j \leqslant d_\text{c}$), it is possible to capture both the shear strength $\sigma_\text{m,N}$ and the surface energy $\gamma$ of the target material. In this lower range of particles' sizes, $\delta_\text{f}$ has a constant value. However, the particles cannot be given an arbitrary small size. Since $\delta_\text{f}$ represents the size of the neighborhood of a particle (see Figure~\ref{fig:dem_deltas}), it indicates from how far a particle can feel a force from another particle. In the case where the diameter of a particle becomes smaller than the neighborhood size $\delta_\text{f}$, the particle will be able to `communicate' with others located further than its closest neighbors, increasing a lot the computational cost of the simulation. Therefore, it is reasonable to keep the particles' sizes over a minimum value of
\begin{equation}
	d_\text{min} = \frac{4\gamma}{\sigma_\text{m,N}} \label{eq:dmin}
\end{equation}
(which is the maximum value of $\delta_\text{f}$). Note that for physically realistic target properties, $d_\text{min}$ can be comparable to the size of atoms, for which the interaction distance is roughly the same as their size. Having the lower limit on the size of the discrete particles bounded by the size of an atom is consistent with our coarse-graining approach. On the other side of the spectrum of particle sizes, when $r_i + r_j > d_\text{c}$, the shear strength $\sigma_\text{m,N}$ and the surface energy $\gamma$ of the target material cannot be both matched at the same time (see Appendix~\ref{apx:force_param} for more details). Depending on the value of the scaling parameter $s$, the tensile strength will be matched to a lower value of
\begin{equation}
	\sigma_\text{m,N}' = \sigma_\text{m,N} \left( \frac{r_i + r_j}{d_\text{c}} \right)^{-s} \,. \label{eq:sms}
\end{equation}
When $s = 0$, the matched tensile strength remains constant, but the matched surface energy is larger than it should be. With $s = 1/2$, the surface energy stays constant, but instead the matched tensile strength decreases with the size of the particles. The decrease of the strength between the particles can be related to the same decrease of the strength of a material when tested with samples of increasing sizes, where the larger samples have a higher chance of containing defects and thus have a lower strength. If one wishes to use this model in both the lower sizes range $d_\text{min} \leqslant r_i + r_j \leqslant d_\text{c}$ and with the larger sizes, the target tensile strength $\sigma_\text{m,N}$ can be taken as the \emph{ideal} tensile strength of the material to model, which is the strength measured at a small scale when no defects are present in the tested sample, and the scaling parameter $s$ can be chosen to represent the desired behavior.

It is worth noting that $d_\text{c}$ \eqref{eq:dc} shares the same expression (ignoring a geometrical multiplication factor) with the \emph{critical length scale} of the target material, established by \textcite{aghababaeiCriticalLengthScale2016}\snspace[0.6em], under which the material has a ductile behavior and exhibits a higher strength\cite{luoSizeDependentBrittletoDuctileTransition2016}.

\section{Validation}\label{sec:validation}

The choice of force parameters given by equations~\eqref{eq:kN} to \eqref{eq:cT} does not ensure that the resulting macroscopic properties will be exactly equaled, since they strongly depend on the coordination number, which in turn is dependent on the volume fraction and the size of the neighborhood of the particles. In order to eliminate these unknowns, we first perform patch tests on systems made of particles arranged in an hexagonal close-packed (HCP) lattice, so that the coordination number is fixed.

We are interested in adhesive wear, which can be regarded as a fracture process at small scale. Ordered systems may have preferential planes for fracture propagation, and we ultimately want to model isotropic materials. Consequently, we also build amorphous systems of particles and perform the same kind of patch tests, this time without full control over the coordination number.

The force parameters are chosen to match the material properties listed in Table~\ref{tab:target_mat_props}. No units are specified, so any coherent system of units can be considered. The value of the target $\nu$ is varied for the lattice tests. From the material properties, we compute the minimum allowed particle size $d_\text{min} = 30$ \eqref{eq:dmin} and the maximum particle size $d_\text{c} = 150$ \eqref{eq:dc} over which the tensile strength and the surface energy cannot be both equaled. For particle sizes greater than $d_\text{c}$, scaling factors of $s = 0$ (constant strength) and $s = 1/2$ (constant surface energy) are investigated.
\begin{table}[H]
	\centering
	\caption{List of target material properties for patch tests}
	\label{tab:target_mat_props}
	\vspace{-0.5em}
	\small
\begin{tabular}{ccccccc}
	\toprule
	$E$ & $\nu$ & $\sigma_\text{m,N}$ & $\sigma_\text{m,T}$ & $\gamma$ & $\eta$ & $\rho$ \\
	\midrule
	1 & 0.15 & 0.2 & 0.1 & 1.5 & 0.95 & 1 \\
	\bottomrule
\end{tabular}

\end{table}

\begin{figure*}
	\centering
	\subfloat[Crystalline system for patch test\label{subfig:patch}]{
		\includegraphics{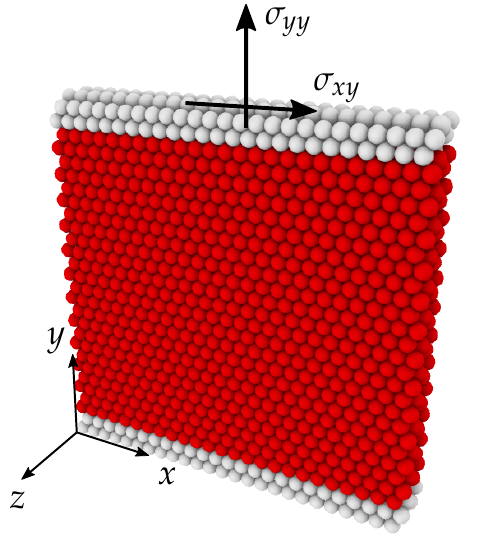}
	}
	\subfloat[Tensile strength test\label{subfig:patch_smN}]{
		\includegraphics{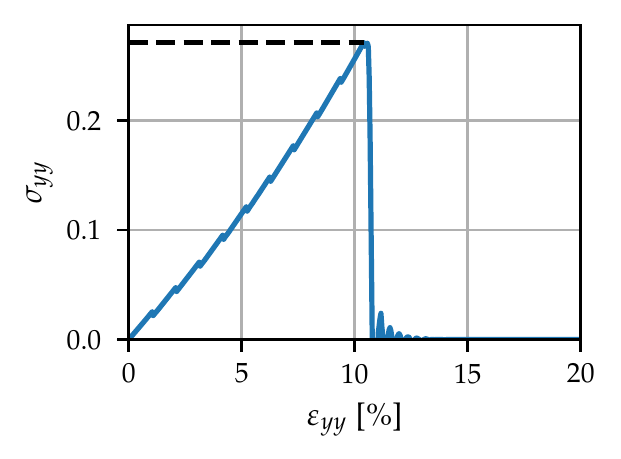}
	}
	\subfloat[Shear strength test\label{subfig:patch_smT}]{
		\includegraphics{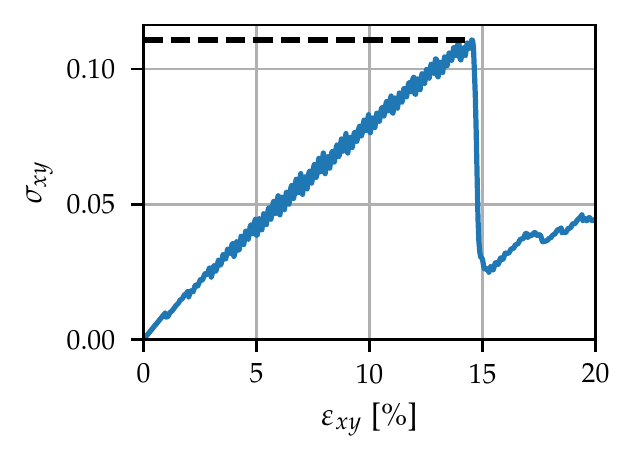}
	}
	\caption[Example of patch test's system and outputs]{Example of patch test's system and outputs. \textbf{\subref{subfig:patch}}~White particles are part of rigid walls. The bottom wall is fixed and a stress or displacement is imposed on the top wall. Boundaries are periodic in the $x$ and $z$ directions. To measure $E$ and $\nu$, a small compressive stress $\sigma_{yy}$ is applied, then $\sigma_{xx}$ and $\varepsilon_{yy}$ are measured. \textbf{\subref{subfig:patch_smN}-\subref{subfig:patch_smT}} A displacement is imposed on the top wall in the appropriate direction to deform the system, and the stresses are monitored. The strengths are defined as the peak measured stresses (dashed lines).}
	\label{fig:patch}
\end{figure*}

The elastic properties are determined by applying a unidirectional compressive load of $0.01$ on a confined sample (see Figure~\ref{fig:patch}\subref{subfig:patch}), up to a deformation of the order of $\varepsilon \sim 0.01$, and measuring the stiffness coefficients $C_{11}$ and $C_{12}$, from which $E$ and $\nu$ are deduced. The loading is performed with a time step of $\Delta t = 0.1 \, \Delta t_\text{c}$ and a global damping of $c = 0.2 \, c_\text{c}$, where $\Delta t_\text{c}$ and $c_\text{c}$ are respectively the critical time step \eqref{eq:Dtc} and the critical damping
\begin{equation}
	c_\text{c} = \sqrt{2k_\text{N}m} \label{eq:cc}
\end{equation}
for a system of two particles (see Appendix~\ref{apx:c_crit}), evaluated for the smallest particle present in the system (leading to the most restrictive time step).

The tensile and shear strengths are determined by deforming the system in the appropriate direction using rigid walls moving at a constant rate of $\dot{\varepsilon} = 10^{-5} / \Delta t$ (with periodic boundary conditions) and measuring the peak stress before failure (see Figures~\ref{fig:patch}\subref{subfig:patch_smN} and \subref{subfig:patch_smT}). The stresses are obtained by computing the average virial stress\cite{moranteStressTensorMolecular2006} inside the deformable part of the system. The simulations are performed with a time step of $\Delta t = 0.1 \, \Delta t_\text{c}$ and a global damping of $c = 0.01 \, c_\text{c}$.

\subsection{Crystalline lattice}

Different particle sizes (diameters) are tested, ranging from $d_0 = 0.6 \, d_\text{c}$ to $d_0 = 76.8 \, d_\text{c}$. The target Poisson's ratio is also varied from $\nu = 0$ to $\nu = 1/4$ (which is the maximum Poisson's ratio acceptable by our model). The size of each tested sample is equal to $L \times L \times W$, with $L = 25\,d_0$ and $W = 3\,d_0$.

The measured Young's moduli and Poisson's ratios match the target properties, with an acceptable deviation (low enough to allow for an easy later adjustment of the force parameters). The error is smallest when $\nu$ is near $0$ and reaches a maximum of $12\%$ when $\nu = 1/4$. The Figures~\ref{fig:lattice_E} and \ref{fig:lattice_nu} in the Appendix depict the actual deviations. Since the organization of particles is always an HCP lattice regardless of the size of the particles $d_0$, the latter has no influence on the measured elastic properties measured at small strain.

\begin{figure}
	\centering
	\includegraphics{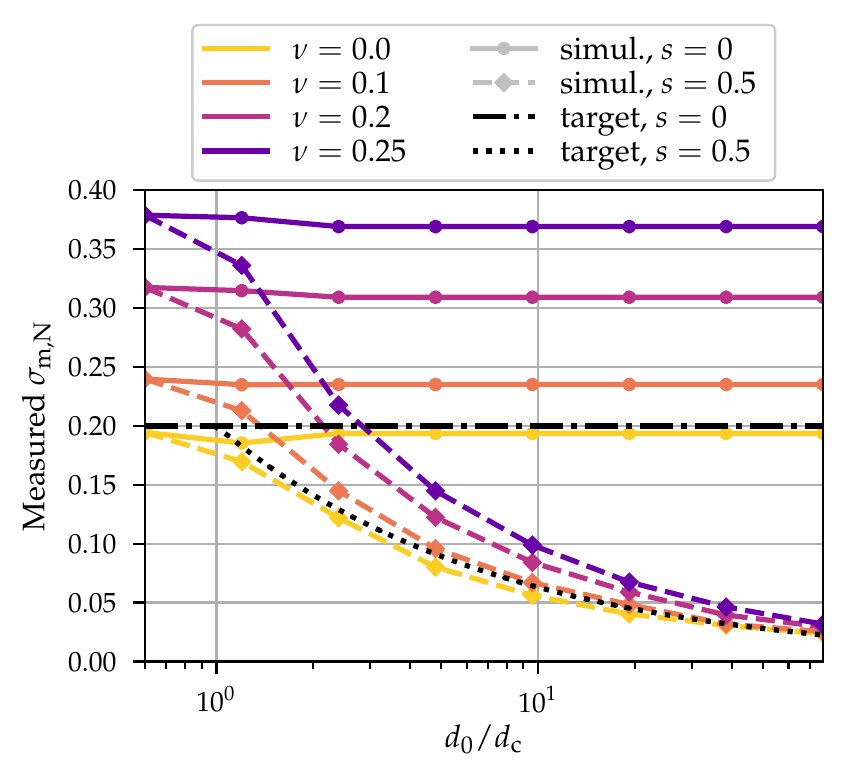}
	\caption[Effect of $d_0$, $s$ and target $\nu$ on the measured tensile strength $\sigma_\text{m,N}$]{Effect of $d_0$, $s$ and target $\nu$ on the measured tensile strength $\sigma_\text{m,N}$. There is one set of curves for $s = 0$ and another one for $s = 0.5$, as indicated by the right half of the legend. The measured $\sigma_\text{m,N}$ matches the target one when $\nu = 0$. For higher values of $\nu$,the strengths are larger than the target ones, but their dependence on $d_0$ remains the same.}
	\label{fig:lattice_sn}
\end{figure}

\begin{figure}
	\centering
	\includegraphics{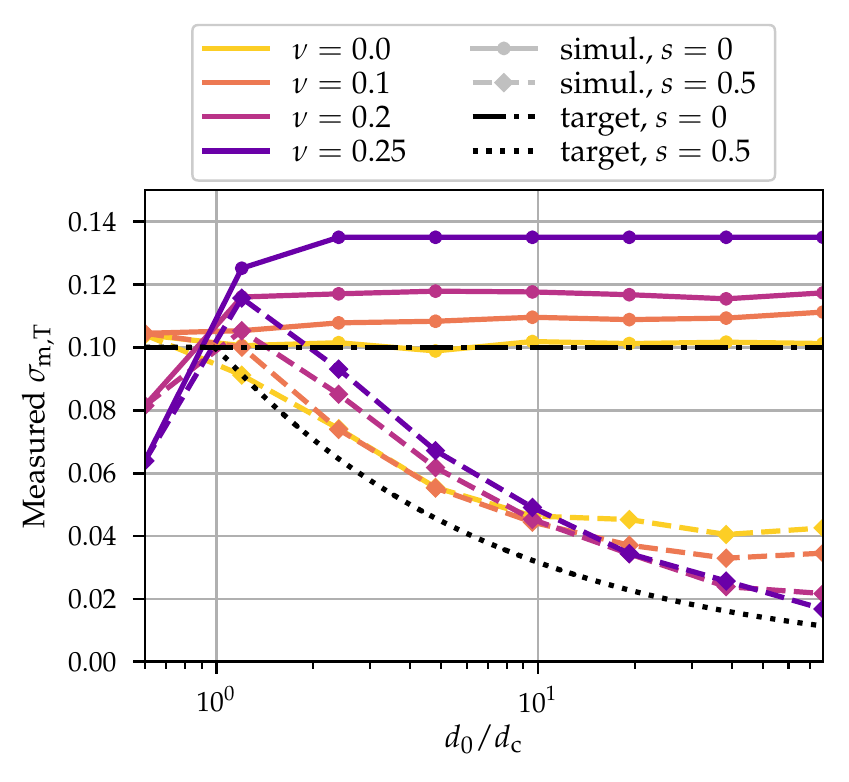}
	\caption[Effect of $d_0$, $s$ and target $\nu$ on the measured shear strength $\sigma_\text{m,T}$]{Effect of $d_0$, $s$ and target $\nu$ on the measured shear strength $\sigma_\text{m,T}$. The behavior of the measured shear strength is mostly the same as the tensile one, except for $d_0 < d_\text{c}$ where the measured strengths are lower than the target ones.}
	\label{fig:lattice_st}
\end{figure}

The measured tensile and shear strengths are shown in Figures~\ref{fig:lattice_sn} and \ref{fig:lattice_st}. They depend on $\nu$, $d_0$, and the scaling parameter $s$. For a scaling of $s=0$, the target tensile and shear strengths (shown by the black dash-dotted curves) are constant with respect to $d_0$, whereas for $s=0.5$, the target strengths (black dotted curves) decrease with respect to $d_0$, as described by \eqref{eq:sms}. The measured strengths match the target ones when the target Poisson's ratio is equal to 0. Otherwise, the strengths are higher than the target ones, while following the same trend with respect to the value of $d_0$. For the shear strength, the measured values are instead lower than the target ones when $d_0 < d_\text{c}$. We did not try to analytically predict the gap between the target and measured strengths. To accurately capture the measured strengths, the values of the target strengths must be adjusted according to the target $\nu$ and to the plots (Figures~\ref{fig:lattice_sn} and \ref{fig:lattice_st}).

\subsection{Amorphous sample}

\paragraph{Particles' size distribution}

In order to obtain an amorphous sample, the particles must have various sizes. Otherwise, particles of identical sizes would arrange into a crystalline lattice or crystalline grains with weaker grain boundaries. We distribute the particles' sizes around a diameter of $d_0$, within the bounds $d_\text{s}$ (the smallest diameter) and $d_\text{l}$ (the largest). The particles' diameters are distributed along a log-normal distribution of mode $d_0$ (most frequent value) and standard deviation $0.2 \, (d_\text{l} - d_\text{s})$. The distribution is truncated between $d_\text{s}$ and $d_\text{l}$. When $d_0$ is at the midpoint between the bounding diameters, the log-normal distribution is similar to a Gaussian distribution. In other cases, this particular distribution allows us to choose a larger $d_\text{l}$ to add a small amount of larger particles inside the sample, while keeping the smallest diameter $d_\text{s}$ and the average diameter $d_0$ the same. The particles are inserted in the system at random positions until they fill it up to a given volume fraction of $0.75$.

\paragraph{Relaxation}

The system of randomly placed particles is relaxed in two phases by simulating it dynamically with a global velocity damping until an equilibrium state is reached. In the first phase, only normal repulsive forces are considered in addition to the global damping forces, allowing the particles to rearrange into a state with no completely overlapped particles. This phase is run for 3000 time steps of $\Delta t = 0.1 \, \Delta t_\text{c}$ with a damping of $c = c_\text{c}$, inside a system with fixed periodic boundaries. For the second phase, the adhesive normal forces are added, and the periodic boundaries are allowed to move in order to adapt to the internal stresses of the system. No tangential forces are considered to avoid the formation of stable holes in the system. This phase is run with a damping of $c = 0.02 \, c_\text{c}$ until all the internal stresses become lower than $10^{-14} E$, where $E$ is the target Young's modulus. Typically, around $20000$ time steps are required for this phase. At the end, the mass of all the particles is adjusted such that the density of the whole system matches the target one. The final volume fraction is likely to change during the relaxation process, along with the movement of the system boundaries.

\paragraph{Patch tests results}

\begin{figure}
	\centering
	\includegraphics{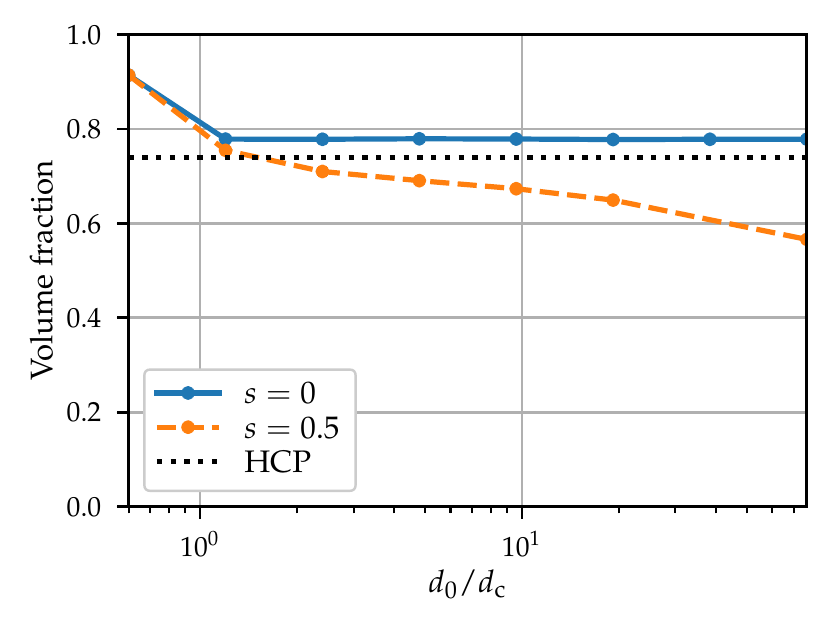}
	\caption[Effect of $d_0$ and $s$ on the relaxed volume fraction]{Effect of $d_0$ and $s$ on the relaxed volume fraction. The volume fraction of HCP lattice systems is shown for comparison. The relaxed volume fraction is independent of $d_0$ when the target shear strength is kept constant ($s = 0$). When the target shear strength decreases with $d_0$ ($s = 0.5$), the relaxed volume fraction also decreases.}
	\label{fig:amorphous_one_vol}
\end{figure}

We test samples of size $L \times L \times W$, with $L = 100\,d_0$, $W = 3\,d_0$, and $d_0$ ranging from $d_0 = 0.6 \, d_\text{c}$ to $d_0 = 76.8 \, d_\text{c}$. The bounds of the particles' size distribution are chosen as $d_\text{s} = 0.75\,d_0$ and $d_\text{l} = 1.25\,d_0$. For each relaxed system, the final volume fraction is measured (see Figure~\ref{fig:amorphous_one_vol}). The scaling parameter $s$ starts to play a role whenever $d_0 > d_c$. When $s = 0$, the ratio between the interaction distance $\delta_\text{f}$ and the particles' diameter $d_0$ remains constant, so the particles can get organized in the same fashion regardless of $d_0$. However, when $s = 0.5$, this same ratio gets smaller when the particles are larger. Less interaction are allowed between the particles, resulting in a system being less densely packed.

\begin{figure}
	\centering
	\includegraphics{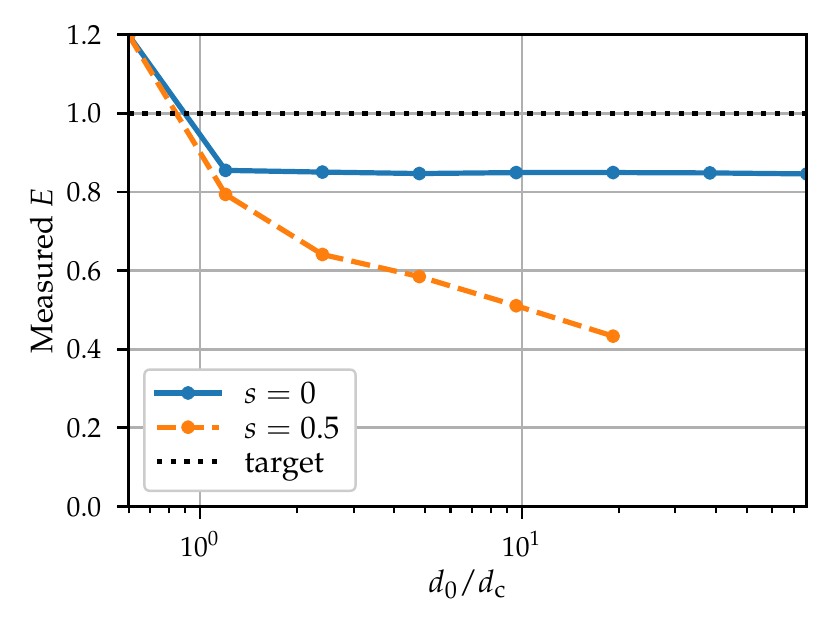}
	\caption[Effect of $d_0$ and $s$ on the measured $E$]{Effect of $d_0$ and $s$ on the measured $E$. The measured Young's modulus follows the same trend as the volume fraction (Figure~\ref{fig:amorphous_one_vol}). For $d_0 > 20 \, d_\text{c}$, the system fails under the compression used to measure the elastic properties.}
	\label{fig:amorphous_one_E}
\end{figure}

The decreased volume fraction has a direct impact on the coordination number of each particles, and thus on the macroscopic elastic properties of the system, as shown by the drastic effect of $d_0$ on the measured Young's modulus when $s = 0.5$ (Figure~\ref{fig:amorphous_one_E}). As $d_0$ increases, the distance of interaction becomes comparatively smaller, resulting in less links between particles and a more fragile network. When $d_0 > 20 \, d_c$, the system no longer resists the compression stress of $0.01$ imposed to measure the elasticity parameters, which is why no values are reported beyond this value of $d_0$. The measured Poisson's ratio stays constant at $\nu \approx 0.20$ regardless of $s$ (see Figure~\ref{fig:amorphous_one_nu} in Appendix), which is higher than the target value ($\nu = 0.15$).

Finally, the measured strengths follow the correct trends with respect to $d_0$ and $s$ (see Figures~\ref{fig:amorphous_one_sn} and \ref{fig:amorphous_one_st} in Appendix). However, the tensile strengths are reduced to $50\%$ of the target one, and the shear strengths are at $56\%$. This is likely due to the particles not being in direct contact with their neighbors ($\delta_\text{N} = 0$), therefore not benefiting from the whole adhesive range $0 \leqslant \delta_\text{N} \leqslant \delta_\text{f}$, resulting in a decreased strength in both normal and tangential directions. This phenomenon is not present in the lattice systems.

\subsection{Discretization}\label{sec:discr}

We saw that when the scaling factor $s$ is not equal to 0, the measured mechanical properties of simulated systems are significantly affected by $d_0$. If one desires to simulate multiple samples having the same size but different discretizations by varying $d_0$, the obtained samples will have different elastic properties because of $d_0$. To mitigate this effect, the distribution of the particles' sizes can be adapted by having the largest bounding diameter $d_\text{l}$ constant throughout all samples (\emph{i.e.} $d_\text{l} = 1.25 \max (d_0)$). The larger particles act like defects in the samples, keeping the strengths as low as when all particles are large. The effect of keeping a constant $d_\text{l}$ for multiple discretizations is presented in detail in Appendix~\ref{apx:discr}. This method is shown to work well to harmonize the strengths across multiple samples of the same size when $s = 0.5$.

\subsection{Calibration}

As already known\cite{chengElasticWavePropagation2020,hentzIdentificationValidationDiscrete2004} and witnessed once more here, obtaining the correct continuum behavior with DEM is a challenging task. Nevertheless, we have shown that our estimates for the force parameters (equations~\eqref{eq:kN} to \eqref{eq:cT}) result in measured elastic and fracture properties being approximately at the target value, especially for a scaling factor of $s = 0$. The force parameters can be further adjusted if necessary, after running the relevant patch tests.

\section{Application: nanoscale adhesive wear}\label{sec:wear}

Molecular dynamics (MD) simulations of adhesive wear have been performed\cite{aghababaeiCriticalLengthScale2016} by modelling two surfaces being in contact at a single junction of a given size and moving in a shear motion relative to each other. In accordance with theoretical predictions\cite{rabinowiczEffectSizeLooseness1958}, it was shown that junctions smaller than a critical size $d^*$ are subjected to plastic smoothening, while junction larger than $d^*$ can detach and form a wear particle. The critical size $d^*$ at first order only depends on material parameters, and defines the boundary between ductile and fragile behaviors in a material.

We use our coarse-grained model to perform the same kind of nanoscale adhesive wear simulation and see if we can reproduce both the ductile and the fragile behaviors for a given material. We choose to model amorphous silica (\ce{SiO2}), which has the material properties listed in Table~\ref{tab:sio2_params}. The coarse-grained model was implemented in LAMMPS\cite{plimptonFastParallelAlgorithms1995}.

\begin{table}[H]
	\centering
	\caption[Amorphous silica properties]{Amorphous silica properties. The ideal (atomic scale) tensile strength $\sigma_\text{m,N}$ is from \textcite{luoSizeDependentBrittletoDuctileTransition2016} and the shear strength is estimated from the tensile one. The restitution coefficient $\eta$ is arbitrarily chosen.}
	\label{tab:sio2_params}
	\vspace{-0.5em}
	\small
\setlength{\tabcolsep}{4pt} 
\begin{tabular}{ccccccc}
	\toprule
	$E$ & $\nu$ & $\sigma_\text{m,N}$ & $\sigma_\text{m,T}$ & $\gamma$ & $\eta$ & $\rho$ \\
	\midrule
	\SI{73}{GPa} & 0.17 & \SI{16}{GPa} & \SI{9}{GPa} & \SI{1.5}{N/m} & 0.9 & \SI{2200}{kg/m^3} \\
	\bottomrule
\end{tabular}

\end{table}

From the material parameters, we compute the critical length scale from the expression of \textcite{aghababaeiCriticalLengthScale2016} for the geometrical configuration we will use:
\begin{equation}\label{eq:dstar}
	d^* \approx \frac{32 \gamma G}{\sigma_\text{m,T}^2} = \SI{18}{nm} \,.
\end{equation}

\begin{figure}
	\centering
	\subfloat[$D = \SI{10}{nm}$, initial\label{subfig:jun-d0_1.5-d_10-ini}]{
		\includegraphics{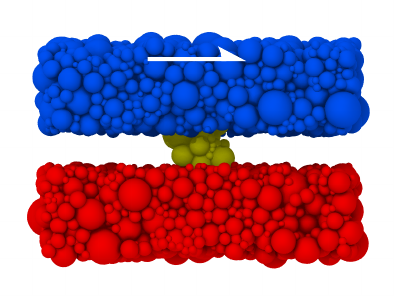}
	}
	\subfloat[$D = \SI{10}{nm}$, \SI{36}{nm} of sliding\label{subfig:jun-d0_1.5-d_10-fin}]{
		\includegraphics{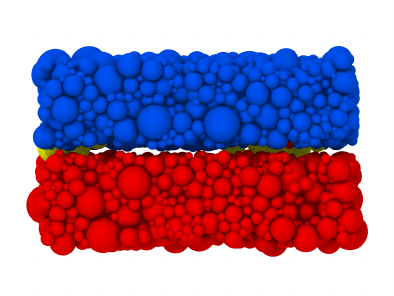}
	}\\
	\subfloat[$D = \SI{20}{nm}$, initial\label{subfig:jun-d0_1.5-d_20-ini}]{
		\includegraphics{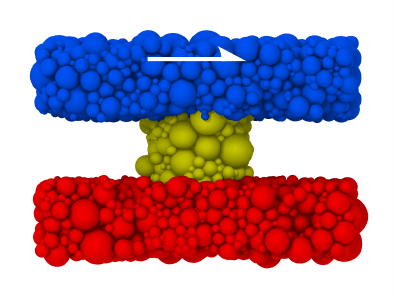}
	}
	\subfloat[$D = \SI{20}{nm}$, \SI{32}{nm} of sliding\label{subfig:jun-d0_1.5-d_20-fin}]{
		\includegraphics{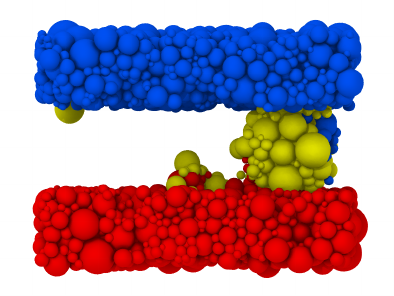}
	}
	\caption[Sheared junctions with $d_0 = \SI{1.5}{nm}$]{Sheared junctions with $d_0 = \SI{1.5}{nm}$. The colors of the particles indicate to which body they initially belong to. The bottom surface is fixed and the top one is dragged from left to right. The system length is \SI{60}{nm} with periodic boundary conditions. \textbf{\subref{subfig:jun-d0_1.5-d_10-ini}-\subref{subfig:jun-d0_1.5-d_10-fin}} The smaller junction deforms plastically and gets squished. \textbf{\subref{subfig:jun-d0_1.5-d_20-ini}-\subref{subfig:jun-d0_1.5-d_20-fin}} The larger junction detaches into a rolling wear particle. The two mechanisms observed with MD simulations are remarkably recovered with the coarse-graining method. Videos for each case are available as supplementary material.}
	\label{fig:jun-d0_1.5}
\end{figure}

We simulate systems of $60 \times 40 \times 40$ \si{nm^3} made of two solids linked by a cylindrical junction of diameter $D = \SI{10}{nm}$ or \SI{20}{nm} and of height $H = D/2$ (see Figures~\ref{fig:jun-d0_1.5}\subref{subfig:jun-d0_1.5-d_10-ini} and \subref{subfig:jun-d0_1.5-d_20-ini}). The two values of $D$ are chosen to have one smaller than $d^*$ and the other one larger.

From the material properties of \ce{SiO2}, we compute the minimum allowed DEM particle size $d_\text{min} = \SI{0.37}{nm}$ \eqref{eq:dmin} and the critical particle size $d_\text{c} = \SI{1.7}{nm}$ \eqref{eq:dc}. From those, we chose to use DEM particles of size $d_0 = \SI{1.5}{nm}$, \SI{3}{nm} and \SI{6}{nm}. For comparison, the bond lengths between atoms in silica are\cite{vashishtaInteractionPotentialSiO21990}{} \ce{Si-O}: \SI{0.16}{nm}, \ce{O-O}: \SI{0.26}{nm} and \ce{Si-Si}: \SI{0.31}{nm}. DEM particles are therefore at least 10 times larger than atoms. For each average particle size $d_0$, we take the bounds of the particles' sizes distribution as $d_\text{s} = 0.75\,d_0$ and $d_\text{l}$ fixed to $d_\text{l} = \SI{7.5}{nm}$, so that every system has the same mechanical properties regardless of $d_0$. We chose the scaling parameter $s = 0.5$ for the dependence of strength on $d_0$.

The amorphous systems are created and relaxed using the same procedure as for the validation tests, resulting in boxes fully filled with particles. The systems are then carved by removing particles to obtain the desired shapes (two surfaces with one cylindrical junction). Two rigid walls of width equal to $1.5\,d_0$ are used to impose a shear motion on the systems, with the bottom one remaining fixed and to top one moving with a constant shear velocity of \SI{10}{m/s}, which is sufficiently small compared to the pressure wave velocity in the medium $c = \sqrt{E/\rho} = \SI{5760}{m/s}$. A constant normal load of \SI{100}{MPa} is applied on the top wall to prevent it from drifting apart, but similar results are obtained with smaller normal loads. The time step is $\Delta t = 0.1 \, \Delta t$.

The results of the simulations are shown in Figure~\ref{fig:jun-d0_1.5} for the finest discretization ($d_0 = \SI{1.5}{nm}$) and in Figure~\ref{fig:jun-d0_6} for the coarsest ($d_0 = \SI{6}{nm}$). The results for $d_0 = \SI{3}{nm}$ can be found in the Appendix (Figure~\ref{fig:jun-d0_3}). All visualizations were rendered using OVITO\cite{stukowskiVisualizationAnalysisAtomistic2009}. For both levels of discretization, the same behaviors emerge. The small junction ($D = \SI{10}{nm}$) gets deformed plastically and squished under the imposed shear, because it is smaller than the critical size $d^*$ of the material. In turn, the large junction ($D = \SI{20}{nm}$), which is larger than the critical $d^*$\snspace, is detached (by fracture) from the surfaces and starts rolling. The coarse-grained DEM approach is able to reproduce both the ductile and brittle behaviors of the simulated material. From the simulations, we deduce that the critical size of the material is in the bounds $\SI{10}{nm} < d^* < \SI{20}{nm}$, which is consistent with the theoretical estimate \eqref{eq:dstar}.

\begin{figure}
	\centering
	\subfloat[$D = \SI{10}{nm}$, initial]{
		\includegraphics{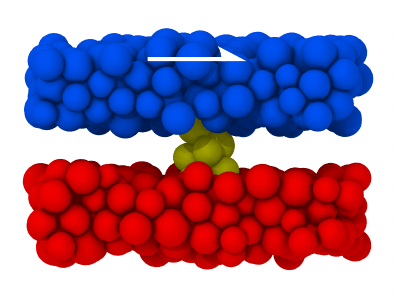}
	}
	\subfloat[$D = \SI{10}{nm}$, \SI{80}{nm} of sliding]{
		\includegraphics{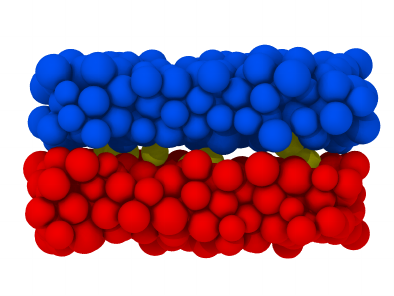}
	}\\
	\subfloat[$D = \SI{20}{nm}$, initial]{
		\includegraphics{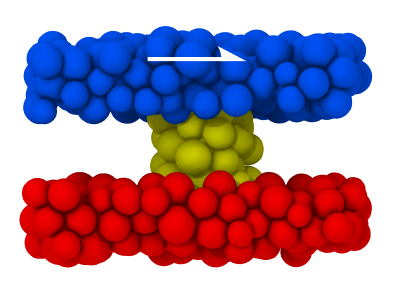}
	}
	\subfloat[$D = \SI{20}{nm}$, \SI{32}{nm} of sliding]{
		\includegraphics{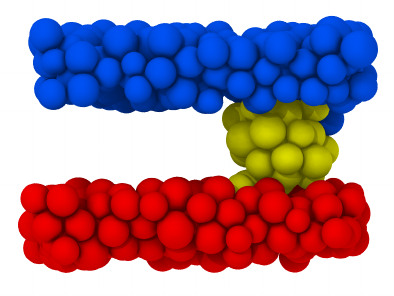}
	}
	\caption[Sheared junctions with $d_0 = \SI{6}{nm}$]{Sheared junctions with $d_0 = \SI{6}{nm}$. The observed behaviors are the same as with the finer discretization of $d_0 = \SI{1.5}{nm}$ (Figure~\ref{fig:jun-d0_1.5}). Videos for each case are available as supplementary material.}
	\label{fig:jun-d0_6}
\end{figure}

We can assume that the ductile behavior $D < d^*$ can only be observed if the DEM particles are sufficiently smaller than $d^*$\snspace, which is the case with all our discretizations. Taking larger particles would result in loosing the ability to model the ductile behavior.

Our method is successfully able to reproduce results that are obtained using MD, while having to simulate less particles and with a larger time step. In addition, silica is a relatively complex and costly material to simulate in MD. For example, the potential of \textcite{vashishtaInteractionPotentialSiO21990} can be used, taking into account 3-body interactions to accurately simulate the bounds between \ce{SiO2} atoms. The Table~\ref{tab:md_dem} compares the estimated computational cost of MD and coarse-grained DEM simulations to perform a simulation equivalent in size and duration to the sheared junction of $D = \SI{20}{nm}$. The computational time for MD simulations is estimated by scaling the time needed to simulate a smaller system on a shorter period of time. The coarse-grained DEM simulations show a definite advantage. However, it should be clear to the reader that the coarse-grained approach results in losing atomistic details (such as three-body interactions and presence of two types of atoms), and that we only aimed to capture rough material properties, in particular for the ductile to brittle transition.

\begin{table}[H]
	\centering
	\caption[Estimated time for a simulation of \SI{100000}{nm^3}\snspace, \SI{10}{ns}, on 28 \SI{2.6}{GHz} CPUs]{Estimated time for a simulation of \SI{100000}{nm^3}\snspace, \SI{10}{ns}, on 28 \SI{2.6}{GHz} CPUs. $N$ is the number of atoms/particles. The simulations are equivalent in size and duration to the sheared junction of $D = \SI{20}{nm}$.}
	\label{tab:md_dem}
	\vspace{-0.5em}
	\small
\begin{tabular}{rrrrr}
	\toprule
	& $d_0$ & $N$ & $\Delta t$ & time \\
	\midrule
	MD & $\approx \SI{0.2}{nm}$ & \SI{7800000}{} & \SI{1}{fs} & 530 days \\
	DEM & \SI{1.5}{nm} & \SI{6000}{} & \SI{20}{fs} & 4\ min \\
	DEM & \SI{3}{nm}   & \SI{2600}{} & \SI{40}{fs} & 2\ min \\
	DEM & \SI{6}{nm}   & \SI{600}{}  & \SI{80}{fs} & 25\ s \\
	\bottomrule
\end{tabular}

\end{table}

\section{Conclusion}

We formulated a pair force to be used with the discrete element method, featuring a reversible cohesive part mimicking the simplest pair potentials used in molecular dynamics. We derived expressions for the parameters of the pair force to match the elastic and fracture properties of a chosen material and showed that the calibration process can be greatly helped by using these expressions. Both crystalline and amorphous solids can be modeled. Finally, we showed that our model can be used to perform coarse-grained simulation of adhesive wear at the scale of asperities, with particles having a diameter 10 times larger than the atoms they replace, and with a computational cost reduced by at least 5 orders of magnitude. This method can be used to perform simulations at a scale inaccessible to molecular dynamics, for example involving the evolution of rough surfaces and third-body elements at a tribological interface.

\section*{Supplementary material}

Supplementary material associated with this article can be found along its online version.

\printbibliography
\balance
\clearpage
\small
\appendix
\renewcommand\thefigure{\thesection.\arabic{figure}}
\renewcommand\thetable{\thesection.\arabic{table}}
\renewcommand\theequation{\thesection.\arabic{equation}}
\setcounter{figure}{0}
\setcounter{table}{0}
\setcounter{equation}{0}

\section{Appendix}

\subsection{Derivation of critical time step and damping}

Let us consider a system of two particles of radii $r_i$ and $r_j$, close enough so that they only interact in the linear Hookean regime with a stiffness $k_\text{N}$ and a velocity damping $c_\text{N}$. We only consider normal relative motion, so that the system can be reduced to only one dimension. The dynamical equations of the system are:
\begin{subequations}
\begin{align}
	m_i \ddot{x}_i + c_\text{N} (\dot{x}_i - \dot{x}_j) + k_\text{N} (x_i - x_j) &= 0 \,, \\
	m_j \ddot{x}_j + c_\text{N} (\dot{x}_j - \dot{x}_i) + k_\text{N} (x_j - x_i) &= 0 \,.
\end{align}
\end{subequations}
Both equations can be combined into one by subtracting one to the other and taking $x = x_i - x_j$ as variable:
\begin{equation}\label{eq:dyn_two_particles}
	m_\text{eff} \ddot{x} + c_\text{N} \dot{x} + k_\text{N} x = 0 \,,
\end{equation}
where $m_\text{eff}$ is the effective mass of the system:
\begin{equation}
	m_\text{eff} = \frac{m_i m_j}{m_i + m_j} \,.
\end{equation}

\subsubsection{Critical time step}\label{apx:dt_crit}

In DEM simulations, it is usual to choose a time step proportional to $\sqrt{m/k_\text{N}}$, with a safety factor that ensures stability\cite{burnsCriticalTimeStep2019}. The exact expression of the critical time step guarantying stability was derived for the central difference scheme\cite{osullivanSelectingSuitableTime2004}. For completeness, we derive it for the symplectic Euler scheme we are using.

To compute the critical time step of the system, numerically integrated using equations \eqref{eq:euler}, we can drop the velocity damping force, so that the remaining force is simply $F = -k_\text{N} x$. The one dimensional integration scheme becomes
\begin{subequations}
\begin{align}
	v_{n + 1} &= v_n - \frac{k_\text{N} x_n}{m_\text{eff}} \, \Delta t \,, \\
	x_{n + 1} &= x_n + v_{n + 1} \, \Delta t \,.
\end{align}
\end{subequations}
Replacing $v_{n+1}$ in the expression of $x_{n+1}$, we obtain a fully explicit scheme:
\begin{subequations} \label{eq:scheme}
\begin{align}
	v_{n + 1} &= v_n - \frac{k_\text{N}}{m_\text{eff}} \, \Delta t \, x_n \,, \\
	x_{n + 1} &= \left(1 - \frac{k_\text{N}}{m_\text{eff}}\Delta t^2 \right) x_n + \Delta t \, v_n \,.
\end{align}
\end{subequations}
To check for the stability of the system, we can monitor the total energy of the system
\begin{equation}
	E_n = \frac{1}{2}k_\text{N}x_n^2 + \frac{1}{2}m_\text{eff}v_n^2
\end{equation}
and make sure that it does not grow unbounded. The expression of the energy prompts us to perform the substitutions
\begin{align}
	\hat{x}_n &= \sqrt{\frac{k_\text{N}}{2E_0}} x_n \\
	\text{and}\quad \hat{v}_n &= \sqrt{\frac{m_\text{eff}}{2E_0}} v_n
\end{align}
for the expression for the initial energy (at $n=0$) to become
\begin{equation}
	1 = \hat{x}_0^2 + \hat{v}_0^2 \,. \label{eq:Enorm}
\end{equation}
Using these substitutions and
\begin{equation}
	\hat{\Delta t} = \sqrt{\frac{k_\text{N}}{m_\text{eff}}} \Delta t \,,
\end{equation}
the integration scheme \eqref{eq:scheme} becomes
\begin{subequations}
\begin{align}
	\hat{v}_{n + 1} &= \hat{v}_n - \hat{\Delta t} \, \hat{x}_n \,, \\
	\hat{x}_{n + 1} &= (1 - \hat{\Delta t}^2) \hat{x}_n + \hat{\Delta t} \, \hat{v}_n \,,
\end{align}
\end{subequations}
which can be written in matrix form:
\begin{equation}
	\begin{bmatrix}
		\hat{x}_{n+1} \\ \hat{v}_{n+1}
	\end{bmatrix}
	=
	\begin{bmatrix}
		1 - \hat{\Delta t}^2 & \hat{\Delta t} \\
		-\hat{\Delta t} & 1
	\end{bmatrix}
	\begin{bmatrix}
		\hat{x}_n \\ \hat{v}_n
	\end{bmatrix}
	,
\end{equation}
or
\begin{equation} \label{eq:matrix_scheme}
	\bm{p}_{n+1} = A \bm{p}_n \,,
\end{equation}
which in turn can be expressed directly as a function of the initial conditions:
\begin{equation}
	\bm{p}_{n} = A^n \bm{p}_0 \,.
\end{equation}
Equation \eqref{eq:Enorm} tells us that the initial adimensionalized position-velocity vector $\bm{p}_0$ has a norm of 1. For the integration scheme to be stable, we must ensure that the norm of $\bm{p}_n$ is not growing toward infinity under the repeated application of $A$ in \eqref{eq:matrix_scheme}. From the eigendecomposition of $A$, we know that
\begin{equation}
	A^n = Q\Lambda^n Q^{-1} \,,
\end{equation}
where $Q$ is the matrix of the eigenvectors of $A$ and $\Lambda$ is the diagonal matrix with the eigenvalues:
\begin{equation}
	\lambda_{1, 2} = \frac{2 - \hat{\Delta t}^2 \pm \hat{\Delta t}\sqrt{\hat{\Delta t}^2 - 4}}{2} \,.
\end{equation}
For $A^n$ to stay bounded and thus have stability, we must have $\max(|\lambda_1|, |\lambda_2|) \leqslant 1$, which is true when $\hat{\Delta t} \leqslant 2$ (the eigenvalues become complex numbers), or
\begin{equation}
	\Delta t \leqslant 2 \sqrt{\frac{m_\text{eff}}{k_\text{N}}} \,,
\end{equation}
where we have $|\lambda_1| = |\lambda_2| = 1$. Taking $m_i = m_j = m$, we have $m_\text{eff} = \frac{m}{2}$ and the stability condition becomes
\begin{equation}
	\Delta t \leqslant \sqrt{\frac{2m}{k_\text{N}}} \,.
\end{equation}
This is the same stability condition as for the central differences scheme\cite{osullivanSelectingSuitableTime2004}.

\subsubsection{Critical damping}\label{apx:c_crit}

The system of two particles described by \eqref{eq:dyn_two_particles} is a conventional damped harmonic oscillator. Depending on the value of the damping coefficient $c_\text{N}$, the system will either oscillate with a decreasing amplitude (underdamped regime) or slowly decay toward the equilibrium position without oscillating (overdamped regime). Between those two regimes lies the critically damped regime, where the system decays as quickly as possible toward its equilibrium. The corresponding critical damping coefficient is
\begin{equation}
	c_\text{c} = 2 \sqrt{k_\text{N}m_\text{eff}} \,.
\end{equation}

\subsection{Derivation of force parameters}\label{apx:force_param}

Let us consider two particles of indexes $i$ and $j$. They interact via normal forces $F_\text{N}$ and tangential forces $F_\text{T}$. These forces can be converted into stresses by dividing them by effective contact cross sections $A_\text{N}$ and $A_\text{T}$ in the normal and tangential direction, respectively:
\begin{align}
	\sigma_\text{N} &= \frac{F_\text{N}}{A_\text{N}} \,, \\
	\sigma_\text{T} &= \frac{F_\text{T}}{A_\text{T}} \,.
\end{align}
The value of the cross sections will be derived later.

The equilibrium distance between the particles is equal to $r_i + r_j$. From it, we can convert the normal separation $\delta_\text{N}$ into a normal deformation:
\begin{equation}
	\varepsilon_\text{N} = \frac{\delta_\text{N}}{r_i + r_j} \,.
\end{equation}
In the elastic range, we must have
\begin{equation}
	\sigma_\text{N} = E \varepsilon_\text{N}
\end{equation}
where $E$ is the target Young's modulus of the material to model. From this relation, we find the expression for the normal stiffness in the elastic range:
\begin{align}
	k_\text{N} &= \frac{F_\text{N}}{\delta_\text{N}} \\
			   &= \frac{A_\text{N} \sigma_\text{N}}{(r_i + r_j) \varepsilon_\text{N}} \\
			   &= \frac{A_\text{N} E}{(r_i + r_j)} \,. \label{eq:kN_apx}
\end{align}
We give a similar expression to the tangential stiffness:
\begin{equation}
	k_\text{T} = \frac{A_\text{T} E}{(r_i + r_j)} \,. \label{eq:kT_apx}
\end{equation}
In our model, the elastic limit is found from the maximum tensile stress of the material:
\begin{equation}
	\varepsilon_\text{e} = \frac{\sigma_\text{m,N}}{E} \,,
\end{equation}
from which we deduce the elastic limit in term of normal separation:
\begin{equation}\label{eq:de_apx}
	\delta_\text{e} = \frac{(r_i + r_j) \sigma_\text{m,N}}{E} \,.
\end{equation}
The maximum tangential force is obtained directly from the maximum tangential stress:
\begin{equation}
	F_\text{m,T} = A_\text{T} \sigma_\text{m,T} \,.
\end{equation}
We can compute the energy needed to break the bond between the two particles, which is equal to the area under the force-separation curve (Figure~\ref{fig:dem_forces}):
\begin{align}
	U_\gamma &= \frac{1}{2} \delta_\text{f} k_\text{N} \delta_\text{e} \\
			 &= \frac{A_\text{N} \sigma_\text{m,N} \delta_\text{f}}{2} \,.
\end{align}
This energy can be linked to the surface energy of the material (two surfaces of area $A_\text{N}$ are created during fracture):
\begin{equation}
	\gamma = \frac{U_\gamma}{2 A_\text{N}} \,.
\end{equation}
We obtain the expression for the fracture separation distance:
\begin{equation}\label{eq:df_apx}
	\delta_\text{f} = \frac{4 \gamma}{\sigma_\text{m,N}} \,.
\end{equation}

To compute the expressions for the damping coefficients, we can express the dynamical equation of the system as in \eqref{eq:dyn_two_particles}.
One solution to this equation is the motion
\begin{align}
	x &= \frac{v_0}{\omega} e^{-\frac{c_\text{N}}{2 m_\text{eff}} t} \sin (\omega t) \,, \\
	\dot{x} &= -\frac{c_\text{N} v_0}{2 \omega m_\text{eff}} e^{-\frac{c_\text{N}}{2 m_\text{eff}} t} \sin (\omega t) + v_0 e^{-\frac{c_\text{N}}{2 m_\text{eff}} t} \cos (\omega t) \,,
\end{align}
where $\omega$ is the natural frequency of the system in the linear elastic range:
\begin{align}
	\omega &= \sqrt{\frac{k_\text{N}}{m_\text{eff}}} \sqrt{1 - \frac{c_\text{N}^2}{4k_\text{N}m_\text{eff}}} \\
		   &\approx \sqrt{\frac{k_\text{N}}{m_\text{eff}}} 
\end{align}
when $c_\text{N}$ is small. This particular solution is the motion of an impact between two particles happening at a time $t = 0$ with a relative velocity of $v_0$ (neglecting the cohesive range, when the particles are not touching). The duration of the impact is approximately $t_\text{f} = \pi / \omega$ when $c_\text{N}$ is small (half of a period of oscillation). The final relative velocity after impact is therefore
\begin{align}
	v_\text{f} &= \dot{x}(t_\text{f}) \\
	           &= v_0 e^{-\frac{\pi c_\text{N}}{2 \omega m_\text{eff}}} \,.
\end{align}
The restitution coefficient is defined as the ratio between the initial and the final velocity:
\begin{align}
	\eta &= \frac{v_0}{v_\text{f}} \\
		 &= e^{-\frac{\pi c_\text{N}}{2 \omega m_\text{eff}}} \\
		 &\approx 1 - \frac{\pi c_\text{N}}{2 \omega m_\text{eff}} \\
		 &\approx 1 - \frac{\pi c_\text{N}}{2 \sqrt{k_\text{N} m_\text{eff}}} \,,
\end{align}
from which we obtain the expression for the damping coefficient
\begin{equation}
	c_\text{N} \approx \frac{2 (1 - \eta)}{\pi} \sqrt{k_\text{N} m_\text{eff}} \,,
\end{equation}
which can also be expressed as a fraction of the critical damping:
\begin{equation}
	c_\text{N} \approx \frac{1 - \eta}{\pi} c_\text{c} \,.
\end{equation}
We give a similar expression to the tangential damping coefficient:
\begin{equation}
	c_\text{T} \approx \frac{2 (1 - \eta)}{\pi} \sqrt{k_\text{T} m_\text{eff}} \,.
\end{equation}

\subsubsection{Effect of particle size}

\begin{figure}[H]
	\centering
	\includegraphics{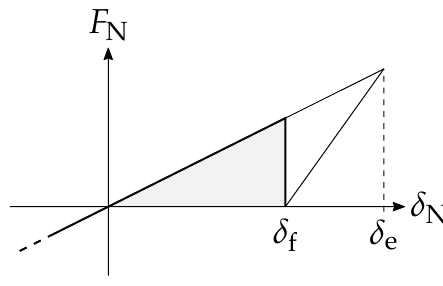}
	\caption{Normal force when $r_i + r_j > d_\text{c}$}
	\label{fig:dem_delta_f_bad}
\end{figure}

Almost all the force parameters depend on the size of the particles, except for the fracture separation distance $\delta_\text{f}$ \eqref{eq:df_apx}. In particular, we can focus on $\delta_\text{e}$ \eqref{eq:de_apx} and $\delta_\text{f}$, which are represented in Figure~\ref{fig:dem_forces}. Since $\delta_\text{e}$ increases with the size of the particles while $\delta_\text{f}$ remains fixed, there is a point at which $\delta_\text{e}$ becomes larger than $\delta_\text{f}$, which happens when $r_i + r_j > d_\text{c}$, where
\begin{equation}
	d_\text{c} = \frac{4\gamma E}{\sigma_\text{m,N}^2}
\end{equation}
is a critical diameter. For large particles, we are in the situation shown in Figure~\ref{fig:dem_delta_f_bad}, with $\delta_\text{e} > \delta_\text{f}$. The plot shows that the elastic limit ($\delta_\text{N} = \delta_\text{e}$) is not reached, and the fracture energy (shaded area) is smaller than expected, meaning that neither the target tensile strength $\sigma_\text{m,N}$ nor the target surface energy $\gamma$ will be matched.

\begin{figure}[H]
	\centering
	\includegraphics{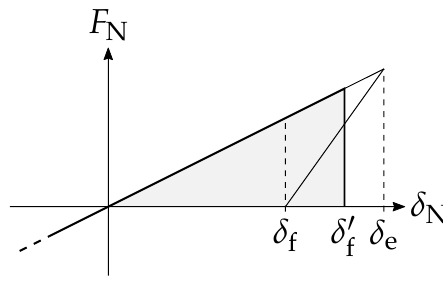}
	\caption{Normal force when $r_i + r_j > d_\text{c}$ with corrected $\delta_\text{f}$}
	\label{fig:dem_delta_f}
\end{figure}

To mitigate this undesirable behavior, we rescale the fracture separation distance $\delta_\text{f}$ to
\begin{equation}
	\delta_\text{f}' = \delta_\text{e} \left( \frac{r_i + r_j}{d_\text{c}} \right)^{-s} \,,
\end{equation}
where $s$ is a scaling factor, resulting in the force plotted in Figure~\ref{fig:dem_delta_f}.

\begin{figure}[H]
	\centering
	\includegraphics{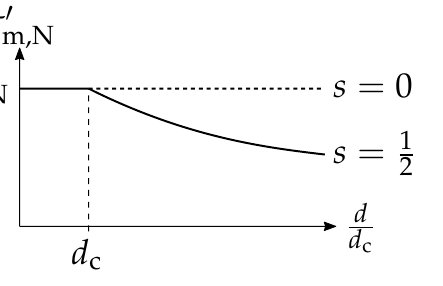}
	\caption{Matched tensile strength with corrected $\delta_\text{f}$}
	\label{fig:dem_sigma_mN}
\end{figure}

By tuning the scaling factor $s$, we can choose to either match the correct elastic limit ($s = 0$, implying $\delta_\text{f} = \delta_\text{e}$), match the target surface energy ($s = 0.5$), or have another behavior. The tensile strength resulting on the choice of $s$ is plotted in Figure~\ref{fig:dem_sigma_mN}.

\subsubsection{Effective cross sections}

The magnitude and the balance between the normal stiffness \eqref{eq:kN_apx} and the tangential stiffness \eqref{eq:kT_apx} are controlled by the effective cross sections $A_\text{N}$ and $A_\text{T}$, and they directly influence the obtained elastic properties (\emph{i.e.} Young's modulus and Poisson's ratio). The effective cross sections must be chosen accordingly.

Following the two-dimensional analysis of \textcite{griffithsModellingElasticContinua2001}, we must express the strain energy stored when deforming a body made of many particles. We start by considering only two particles of the body, both having a radius $r$. The center of one particle is taken as the origin of an arbitrary frame, and the other particle has the spherical coordinates $(2r, \theta, \phi)$ in this frame ($\theta$ is the polar angle, $\phi$ is the azimuthal angle). The deformation of the whole body $\varepsilon_x$ and $\varepsilon_z$ in the $x$ and $z$ directions ($z$ is the zenith direction) directly influence the separation vector $\bm{\delta}$ between the two particles, which can be expressed in the Cartesian frame:
\begin{align}
	\delta_x &= 2r \varepsilon_x \cos\phi \cos\theta \\
	\delta_z &= 2r \varepsilon_z \sin\theta \,,
\end{align}
and in the spherical frame:
\begin{align}
	\delta_r &= \delta_x \cos\phi \cos\theta + \delta_z \sin\theta \\
	\delta_\theta &= -\delta_x \cos\phi \cos\theta + \delta_z \cos\theta \\
	\delta_\phi &= -\delta_x \sin\theta \,.
\end{align}
Note that $\delta_r = \delta_\text{N}$. In the linear elastic range, the strain energy of the single pair of particles is
\begin{equation}
	U_\text{pair} = \frac{1}{2} k_\text{N} \delta_\text{N}^2 + \frac{1}{2} k_\text{T} (\delta_\theta^2 + \delta_\phi^2) \,.
\end{equation}
The total strain energy stored by all possible pairs with neighbors surrounding a single particle is obtained by integrating the strain energy of a single pair:
\begin{equation}
	U = \frac{1}{2} \int_{\phi = 0}^{2\pi} \int_{\theta = -\frac{\pi}{2}}^{\frac{\pi}{2}} U_\text{pair} \cos\theta \, d\theta \, d\phi \,,
\end{equation}
where the leading $1/2$ factor distributes the energy between the single considered particle and its neighborhood. This energy assumes that one particle can be fully surrounded by $4\pi$ neighbors, while the number of neighbors in the most densely packed arrangement of particles (\emph{e.g.} HCP) is 12. We can rescale the strain energy to take this into account:
\begin{equation}
	U' = \frac{12}{4\pi} U \,.
\end{equation}
In an HCP lattice, the particle is surrounded by 12 neighbors positioned at the edges of a cube of side length $2\sqrt{2}r$ (or at the centers of the faces of a rhombic dodecahedron). Each particle can be assigned a piece of the deformable body having a volume of
\begin{equation}
	V = 4\sqrt{2} r^3 \,,
\end{equation}
which is a bit larger than the volume of the spherical particle itself, $4\pi r^3/3$ (the ratio between the two volumes is around 74\%). From the volumetric strain energy, we can make the elastic constants of the granular body appear:
\begin{equation}
	\frac{1}{V} \frac{\partial U'}{\partial \varepsilon_x} = C_{11} \varepsilon_x + C_{12} \varepsilon_z
\end{equation}
Plugging all the expressions of $\bm{\delta}$ into this last equation and identifying the leading factors of $\varepsilon_x$ and $\varepsilon_z$, we obtain the expression of the elastic constants:
\begin{align}
	C_{11} &= \frac{\sqrt{2}}{5r} (3 k_\text{N} + 2 k_\text{T}) \,, \\
	C_{12} &= \frac{\sqrt{2}}{5r} (k_\text{N} - k_\text{T}) \,.
\end{align}
Inverting these, we get the expressions for the stiffness coefficients (expressed directly in term of the Young's modulus and the Poisson's ratio):
\begin{align}
	k_\text{N} &= \frac{\sqrt{2}Er}{2(1 - 2\nu)} \,, \\
	k_\text{T} &= \frac{\sqrt{2}Er (1 - 4\nu)}{2 (1 + \nu) (1 - 2\nu)} \,.
\end{align}
Expressing these in term of the effective cross sections (\eqref{eq:kN_apx} and \eqref{eq:kT_apx}), we finally obtain the expressions for the latter:
\begin{align}
	A_\text{N} &= \sqrt{2} r^2 \frac{1}{1 - 2\nu} \,, \\
	A_\text{T} &= \sqrt{2} r^2 \frac{1 - 4\nu}{(1 - 2\nu)(1 + \nu)} \,.
\end{align}
Hence, all force parameters have been identified. For now, the effective cross sections are defined for particles all having the same radius of $r$.

\subsubsection{Effective particle radius}

In the expressions of the effective cross sections, we can replace the particle radius $r$ by an effective radius $r_\text{eff}$ to take into consideration the two different radii of the pair of particles for which the force is being computed. There are many ways to define the effective radius. We choose
\begin{equation}
	r_\text{eff} = \min (r_i, r_j) \,,
\end{equation}
which results in the macroscopic elastic and strength properties to be controlled by the presence of large particles in the system (as demonstrated in Section~\ref{sec:discr}).

\subsection{Validation plots}

\subsubsection{Crystalline lattice}

\begin{figure}[p]
	\centering
	\includegraphics{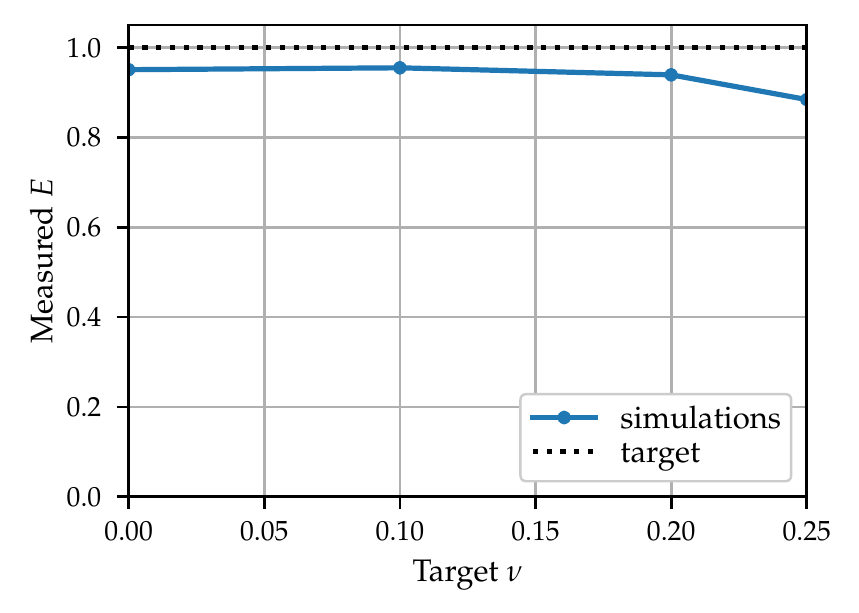}
	\caption{Effect of target $\nu$ on the measured $E$}
	\label{fig:lattice_E}
\end{figure}

\begin{figure}[p]
	\centering
	\includegraphics{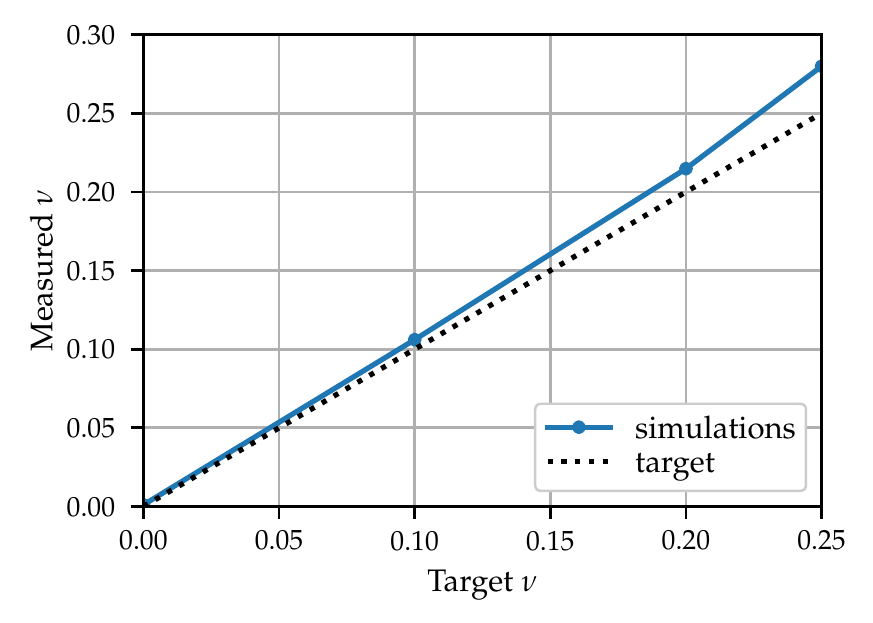}
	\caption{Effect of target $\nu$ on the measured $\nu$}
	\label{fig:lattice_nu}
\end{figure}

The deviations between the target and measured Young's moduli $E$ and Poisson's ratios $\nu$ for crystalline lattice systems are plotted in Figures~\ref{fig:lattice_E} and \ref{fig:lattice_nu} respectively.

\subsubsection{Amorphous sample}

\begin{figure}[p]
	\centering
	\includegraphics{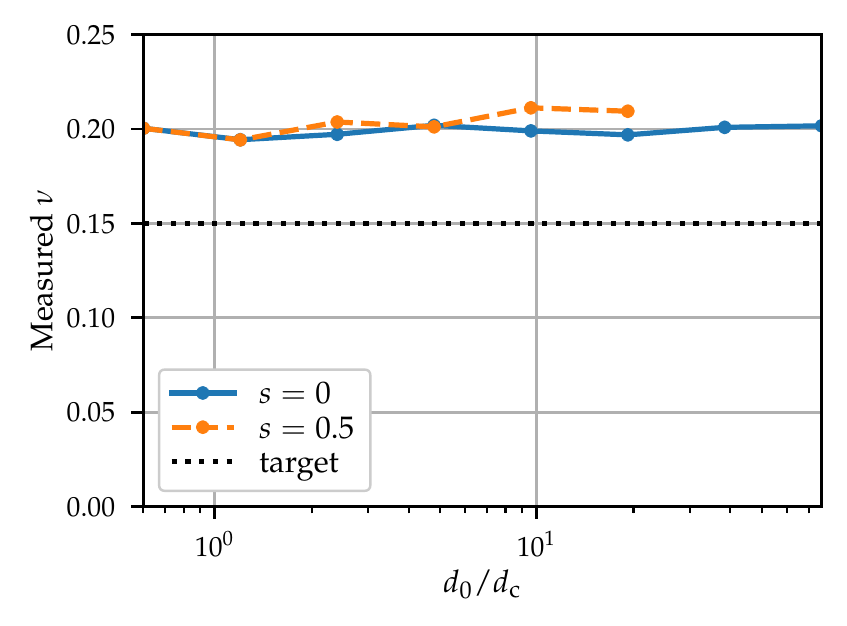}
	\caption{Effect $d_0$ and $s$ on the measured $\nu$}
	\label{fig:amorphous_one_nu}
\end{figure}

\begin{figure}[p]
	\centering
	\includegraphics{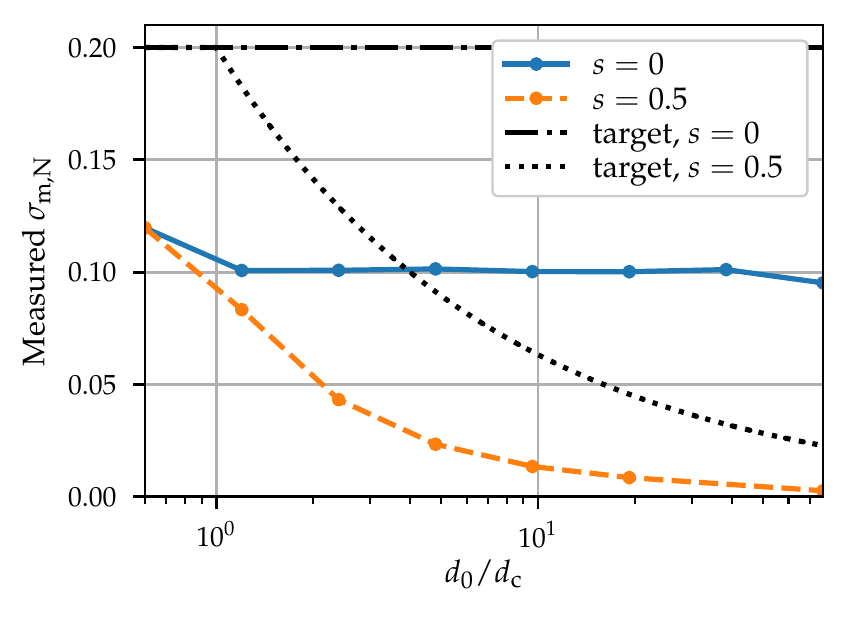}
	\caption{Effect $d_0$ and $s$ on the measured tensile strength $\sigma_\text{m,N}$}
	\label{fig:amorphous_one_sn}
\end{figure}

\begin{figure}[p]
	\centering
	\includegraphics{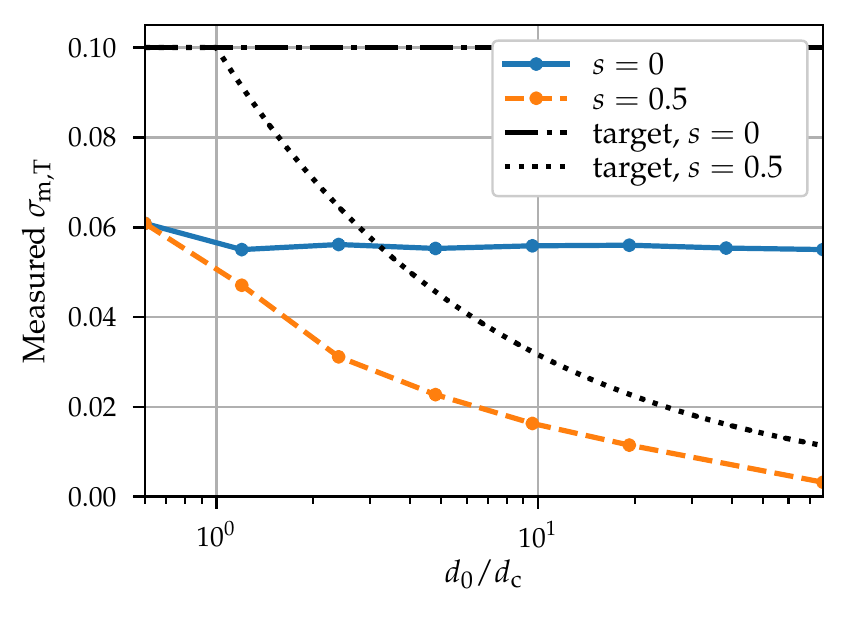}
	\caption{Effect $d_0$ and $s$ on the measured shear strength $\sigma_\text{m,T}$}
	\label{fig:amorphous_one_st}
\end{figure}

The measured Poisson's ratios, tensile strengths and shear strength for amorphous systems are plotted in Figures~\ref{fig:amorphous_one_nu}, \ref{fig:amorphous_one_sn} and \ref{fig:amorphous_one_st}.

\subsection{Discretization}\label{apx:discr}
\nobalance

\begin{figure}[p]
	\centering
	\includegraphics{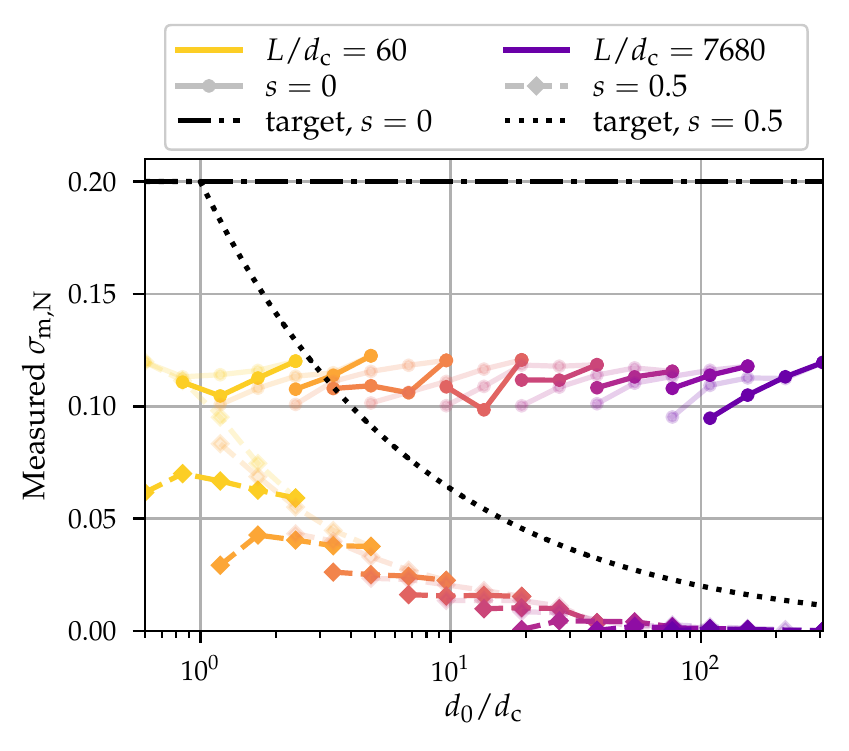}
	\caption[Effect of $d_0$ and $s$ on the measured tensile strength $\sigma_\text{m,T}$ for systems of various sizes and discretizations]{Effect of $d_0$ and $s$ on the measured tensile strength $\sigma_\text{m,T}$ for systems of various sizes and discretizations. Each color corresponds to a fixed system size. The semi-transparent curves correspond to a varying $d_\text{l} = 1.25 \, d_0$, whereas the fully-visible ones have a fixed $d_\text{l} = 0.05 \, L$.}
	\label{fig:amorphous_fixed_sn}
\end{figure}

We test samples of size $L \times L \times W$, with $L$ ranging from $L = 60 \, d_\text{c}$ to $L = 7680 \, d_\text{c}$. The discretization of each system is determined by $d_0$, chosen such that the coarsest systems have $d_0 = L/25$, and the finest have $d_0 = L/100$, while keeping $d_0$ between $0.6 \, d_\text{c}$ and $76.8 \, d_\text{c}$. The smallest bound of the particles' size distribution is chosen as $d_\text{s} = 0.75 \, d_0$, and the largest bound is set to $d_\text{l} = 0.05 \, L$, such that it remains fixed with respect to the system size and is not affected by $d_0$. The thickness of the systems is fixed at $W = 3 \, d_0$. The measured tensile strengths are plotted in Figure~\ref{fig:amorphous_fixed_sn}. The simulations show that, given one system size $L$, having the same largest size particle $d_\text{l}$ for all discretizations results in the systems exhibiting roughly the same tensile strengths.

\subsection{Adhesive wear simulations}

\begin{figure}[p]
	\centering
	\subfloat[$D = \SI{10}{nm}$, initial]{
		\includegraphics{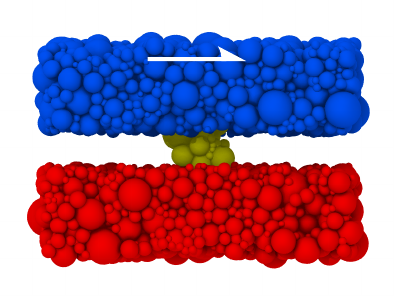}
	}
	\subfloat[$D = \SI{10}{nm}$, after sliding]{ 
		\includegraphics{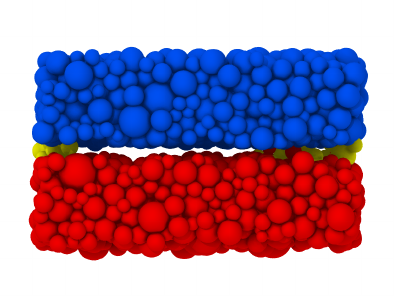}
	}\\
	\subfloat[$D = \SI{20}{nm}$, initial]{
		\includegraphics{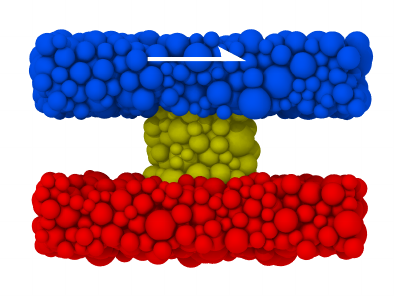}
	}
	\subfloat[$D = \SI{20}{nm}$, after sliding]{ 
		\includegraphics{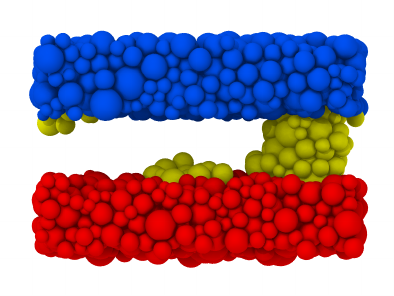}
	}
	\caption[Sheared junctions with $d_0 = \SI{3}{nm}$]{Sheared junctions with $d_0 = \SI{3}{nm}$. The observed behaviors are the same as with the finer and coarser discretizations of $d_0 = \SI{1.5}{nm}$ and $d_0 = \SI{6}{nm}$ (Figures~\ref{fig:jun-d0_1.5} and \ref{fig:jun-d0_6}). Videos for each case are available as supplementary material.}	\label{fig:jun-d0_3}
\end{figure}

The results of the sheared junctions simulations for the intermediate discretization size $d_0 = \SI{3}{nm}$ are shown in Figure~\ref{fig:jun-d0_3}. As seen with the finer (\SI{1.5}{nm}) and coarser (\SI{6}{nm}) discretizations, the small junction gets deformed plastically and squished under the imposed shear, while the large junction is detached from the surfaces and starts rolling. Having the same behaviors emerge whichever the discretization confirms that the material properties are similar.

\end{document}